\documentclass[%
 aip,
 amsmath,amssymb,
 reprint,%
]{revtex4-1}

\usepackage{graphicx}
\usepackage{dcolumn}
\usepackage{bm}
\usepackage{amsmath}
\usepackage{xcolor}
\usepackage[utf8]{inputenc}
\usepackage[T1]{fontenc}
\usepackage{mathptmx}
\usepackage{etoolbox}
\usepackage{comment}
\usepackage{float}

\renewcommand{\Re}{\operatorname{\rm{Re}}}

\makeatletter
\def\@email#1#2{%
 \endgroup
 \patchcmd{\titleblock@produce}
  {\frontmatter@RRAPformat}
  {\frontmatter@RRAPformat{\produce@RRAP{*#1\href{mailto:#2}{#2}}}\frontmatter@RRAPformat}
  {}{}
}%
\makeatother
\begin{document}

\preprint{AIP/123-QED}

\title[Energy exchange between electrons and ions driven by ITG-TEM turbulence]{Energy exchange between electrons and ions driven by ITG-TEM turbulence}
\author{T. Kato}
 \email{kato.tetsuji21@ae.k.u-tokyo.ac.jp}
 \affiliation{Graduate School of Frontier Science, The University of Tokyo, Kashiwa 277-8561, Japan}

\author{H. Sugama}%
\affiliation{Graduate School of Frontier Science, The University of Tokyo, Kashiwa 277-8561, Japan}
\affiliation{National Institute for Fusion Science, Toki 509-5292, Japan}%

\author{T.-H. Watanabe}
\affiliation{Department of Physics, Nagoya University, Nagoya 464-8602, Japan}%


\date{\today}

\begin{abstract}
In this study, the energy exchange between electrons and ions in trapped-electron-mode (TEM) and mixed ion-temperature-gradient (ITG) – TEM turbulence is investigated using gyrokinetic simulations.
The energy exchange in TEM turbulence is primarily composed of the cooling of electrons associated with perpendicular $\nabla B$-curvature drift and the heating of ions moving parallel to magnetic field lines.
TEM turbulence facilitates energy transfer from electrons to ions, which is opposite to the direction observed in ITG turbulence.
In mixed ITG-TEM turbulence, the relative magnitudes of parallel heating and perpendicular cooling for each species determine the overall direction and magnitude of energy exchange.
From the viewpoint of entropy balance, it is further confirmed that energy flows from the species with larger entropy production, caused by particle and heat fluxes, to the other species in ITG-TEM turbulence.
The predictability of turbulent energy exchange in ITG-TEM turbulence by the quasilinear model is examined. 
In addition, an alternative method based on the correlation between energy flux and energy exchange is developed, and its validity is demonstrated.
\end{abstract}

\maketitle

\section{Introduction}

Turbulence in magnetically confined plasmas induces particle flux, heat flux, and energy exchange between electrons and ions\cite{Kato2024, Sugama1996, Sugama2009, Waltz,  Waltz1997, Candy, Manheimer}.
Gyrokinetic theory provides a robust framework for understanding turbulence in magnetized plasmas, and gyrokinetic simulations are widely used to analyse the turbulent transport\cite{Antonsen, CTB, F-C, Pueschel, Dimits, Krommes2012, Garbet2010, Idomura2006, Horton}.
The prediction of turbulent effects is necessary for simulations of global density and temperature profiles.
The quasilinear model is useful for estimating the nonlinear simulation results by linear simulations, which require relatively low computational costs\cite{Casati, Citrin, Citrin2017, Bourdelle, Staebler2021, Staebler2024, Merz, Narita, Toda, Parker, Barnes}.
It is well-established that the global profiles are predominantly determined by a balance between sources of particles and energy, and losses due to turbulent fluxes.
However, the effect of turbulent energy exchange had been considered negligible~\cite{Waltz,  Waltz1997, Candy, Manheimer}.
As a result, while turbulent transport fluxes are commonly incorporated into transport simulations for predicting density and temperature profiles, the contribution of turbulent energy exchange is typically omitted.
Our previous work~\cite{Kato2024}, which investigates energy exchange caused by pure ITG turbulence in tokamak plasmas, has reported that in high temperature or weakly collisional plasmas, such as those expected in future fusion reactors, turbulent energy exchange could exceed collisional one.
Furthermore, it is found that turbulent energy exchange in ITG turbulence can transfer energy from ions to electrons regardless of whether electrons or ions are hotter.
This is significantly different from the collisional energy exchange, which transfers energy from the hotter particle species to the cooler species.
In other words, ITG turbulence can transfer energy from cooler ions to hotter electrons, which appears to contradict the second law of thermodynamics.
However, this seemingly paradoxical phenomenon is explained by the fact that the entropy production associated with ion particle and heat fluxes exceeds that of the electrons,  thereby demonstrating that it does not contradict the second law.
Based on this observation, it is conjectured that energy is generally transferred by turbulence from a particle species with larger entropy production due to particle and heat transport driven by the instability to other species, regardless of which species is hotter.
This study examines the validity of this conjecture in the context of ITG–TEM turbulence.
Since ITG turbulence not only enhances particle and heat transport but also inhibits the energy transfer from alpha-heated electrons to ions, suppression of ITG turbulence is favorable for sustaining the high ion temperature required for fusion reactions.
Thus, transport simulations predicting global temperature profiles should include the effect of turbulent energy exchange for future fusion reactors.
To support this, the feasibility of the quasilinear model for energy exchange due to ITG turbulence in tokamak plasmas has also been demonstrated in our previous work.
Other previous studies~\cite{Manheimer, Waltz1997, Waltz} have examined energy exchange driven by TEM and mixed ITG–TEM turbulence.
In this study, we analyze the entropy balance and energy fluxes as well as the wavenumber spectrum of energy exchange, which have not been addressed in the previous studies, to clarify the physical mechanism of energy exchange in ITG-TEM turbulence.
We also extend our previous work to examine the applicability of the quasilinear model to energy exchange in TEM and mixed ITG–TEM turbulence, by varying the most unstable mode from TEM to ITG.
In addition, we propose an alternative approach that combines the quasilinear model with the correlation between energy fluxes and energy exchange, and demonstrate its effectiveness.
It should be noted that the saturation rule, which is essential for the practical application of quasilinear models, is not addressed here.
Instead, this study focuses on evaluating the validity of a fundamental assumption of the quasilinear model that the ratios of turbulent fluxes and energy exchange to the amplitude of turbulent fluctuations can be approximated by those obtained from linear calculations.
The rest of this paper is organized as follows. 
In Sec.~\ref{sec:2}, the entropy balance equation for the turbulent fluctuation and components of turbulent energy transfer are introduced.
Following the previous paper\cite{Kato2024}, the symbol $Q$ is used to denote the energy exchange while the energy flux is denoted by $\mathcal{E}$ in this paper.
In Sec.~\ref{sec:3}, simulation results of pure TEM and mixed ITG-TEM turbulence by the GKV code\cite{GKV}, which uses a flux tube domain\cite{Beer}, are shown. 
Simulation settings and results of linear simulations are shown in Sec.~\ref{subsec:3A}.
The nonlinear simulation results for pure TEM and mixed ITG-TEM turbulence are presented in Secs.~\ref{subsec:3B} and \ref{subsec:3C}, respectively, and the entropy balance and energy transfer are discussed.
The feasibility of the quasilinear model for energy transfer in ITG-TEM turbulence is investigated in Sec.~\ref{subsec:3D}.
Finally, conclusions and discussion are given in Sec.~\ref{sec:4}.

\section{Theoretical model}
\label{sec:2}
\subsection{Entropy balance equation}
\label{subsec:2A}
We here introduce the balance equation for the entropy associated with the perturbed distribution function.
The perturbed nonadiabatic distribution function with a perpendicular wavenumber $\bm{k}$ for particle species $a=e, i$, denoted as $h_{a\bm{k}}$, is related to the perturbed distribution function $f_{a\bm{k}}$ by the following relation:
\begin{equation}
    f_{a\bm{k}}=-\frac{e_a f_{Ma} }{T_a}\phi_{\bm{k}} +h_{a\bm{k}} e^{-i\bm{k}\cdot \bm{\rho}_{a}},
    \label{eq:distribution function}
\end{equation}
where $\bm{\rho}_{a}=\bm{b}\times\bm{v}/\Omega_a$, $\Omega_a=e_aB/(m_a c)$, $\bm{b}=\bm{B}/B$, $B=|\bm{B}|$ and $c$, $e_a$, $m_a$, $T_a$, $f_{Ma}$, and $\bm{B}$ are the speed of light, charge, mass, temperature, local Maxwellian, and background magnetic field, respectively.
The gyrophase-averaged perturbed potential function $\psi_{a\bm{k}}$ is defined in terms of the perturbed electrostatic potential $\phi_{\bm{k}}$ and the perturbed vector potential $\bm{A}_{\bm{k}}$ as
\begin{equation}
\label{eq: NonAdGK}
    \psi_{a\bm{k}}=J_0\left( k \rho_a \right)\left( \phi_{\bm{k}}-\frac{v_\parallel}{c}A_{\parallel \bm{k}} \right)+J_1\left( k \rho_a \right)\frac{v_\perp}{c}\frac{B_{\parallel \bm{k}}}{k},\\
\end{equation}
where $J_0$ and $J_1$ are the zeroth- and first-order Bessel functions, respectively, $A_{\parallel \bm{k}}=\bm{b}\cdot\bm{A}_{\bm{k}}$, $ B_{\parallel \bm{k}}=i\bm{b}\cdot ( \bm{k}\times\bm{A}_{\bm{k}})$, $\rho_a=v_\perp/\Omega_a$, $k=|\bm{k}|$, $v_\parallel=\bm{v}\cdot\bm{b}$, and $v_\perp=|\bm{v}-v_\parallel \bm{b}|$.
In this study, the background radial electric field $E_r$ is set to zero, since its effect appears only in the Doppler shift in the local gyrokinetic theory and does not fundamentally affect the turbulent transport fluxes and energy exchange.
The background $E\times B$ flow shear effect is also neglected here since it is regarded as smaller by an order of the gyrokinetic expansion parameter $\delta$ than the terms included in the gyrokinetic equation introduced in Ref.~\cite{Kato2024}.
The detailed derivation is explained in Refs.~\cite{Kato2024, Sugama1996, Sugama2009}.
The entropy balance equation in the perpendicular wavenumber space is described as,
\begin{equation}
\frac{\partial}{\partial t}
\left(\delta S_{ha\bm{k}}\right)
=\frac{\Gamma^{\rm turb}_{a\bm{k}}}{L_{pa}}+\frac{q^{\rm turb}_{a\bm{k}}}{T_aL_{Ta}}+\frac{Q^{\rm turb}_{a \bm{k}}}{T_a}+ D_{a\bm{k}}+ N_{a\bm{k}}  
,
\label{eq:EBequation_wavenumber}
\end{equation}
where the perturbed entropy for nonadiabatic distribution function $\delta S_{ha\bm{k}}$, particle flux $\Gamma_{a\bm{k}}^{\rm turb}$, heat flux $q^{\rm turb}_{a\bm{k}}$, energy transfer from the perturbed field to particles $Q^{\rm turb}_{a\bm{k}}$, collisional dissipation $D_{a\bm{k}}$, and entropy transfer by nonlinear interaction $N_{a\bm{k}}$, pressure and temperature gradient scale lengths $L_{pa}$, $L_{Ta}$ are described as,
\begin{eqnarray}
&&\delta S_{ha\bm{k}}=\Big\langle\Big\langle \int d^3v \frac{\left|h_{a\bm{k}} \right|^2}{2f_{Ma}} \Big\rangle\Big\rangle, \label{eq:the perturbed entropy}\\
&&\left[\Gamma^{\rm turb}_{a\bm{k}}, q^{\rm turb}_{a\bm{k}}\right]=\Re\Bigg\langle\Bigg\langle \int d^3v \left[ 1, \frac{m_av^2}{2}-\frac{5T_a}{2}\right]\nonumber\\
&& \hspace{2.5cm}\times h_{a\bm{k}}^*\left(-i\frac{c}{B}\psi_{a\bm{k}}\bm{k}\times\bm{b}\right)\cdot\nabla r \Bigg\rangle\Bigg\rangle, \label{eq:Particle Heat fluxes}\\
&&Q^{\rm turb}_{a\bm{k}}=\Re\Bigg\langle\Bigg\langle \int d^3v e_ah_{a\bm{k}}^*\frac{\partial\psi_{a\bm{k}}}{\partial t} \Bigg\rangle\Bigg\rangle, \label{eq:TurbulentEnergyExchange}\\
&&D_{a\bm{k}}=\Re\Big\langle\Big\langle \int d^3v \frac{ h_{a\bm{k}}^*}{f_{Ma}}C^{GK}_a \Big\rangle\Big\rangle,
\label{eq:CollisionalDissipation}\\
&&N_{a\bm{k}}=\nonumber \\
&&\Re\Bigg\langle\Bigg\langle\int d^3v \frac{c}{Bf_{Ma}} 
\sum_{\bm{k}'+\bm{k}''=\bm{k}}
\left[ \bm{b}\cdot\left( \bm{k}'\times\bm{k}'' \right)\right] 
\psi_{a\bm{k}'}h_{a\bm{k}''}h_{a\bm{k}}^*\Bigg\rangle\Bigg\rangle, \nonumber \\
\label{eq:entropytransfer}\\
&&\left[ L_{pa}^{-1}, L_{Ta}^{-1}\right]=\left[ -\frac{\partial \ln p_a}{\partial r}, -\frac{\partial \ln T_a}{\partial r}\right],
\end{eqnarray}
respectively.
The superscript $*$ and "$\rm Re$" denote the complex conjugate and real parts, respectively, and $p_a$ and $C^{GK}_a$ are the pressure and gyrokinetic collision operator, respectively.
The double angle bracket $\langle \langle \cdots \rangle \rangle$ denotes the double average over the ensemble and flux surface, and $v$ is the absolute value of the velocity $\bm{v}$.
The minor radius $r$ is used as a label of flux surfaces of a toroidal plasma.

Taking the summation of Eq.~(\ref{eq:EBequation_wavenumber}) over wavenumber space, we obtain
\begin{equation}
\frac{\partial}{\partial t}
\left(\delta S_{ha}\right)
=\frac{\Gamma^{\rm turb}_{a}}{L_{pa}}+\frac{q^{\rm turb}_{a}}{T_aL_{Ta}}+\frac{Q^{\rm turb}_{a}}{T_a}+ D_{a},
\label{eq:EBequation_real}
\end{equation}
where $\sum_{\bm{k}}N_{a\bm{k}}=0$.
Under the transport time-scale ordering\cite{Sugama1996, Sugama2009, Hinton1976} $\partial \langle \langle \cdots \rangle \rangle / \partial t = \mathcal{O}(\delta^2)$, the L.H.S. of Eq.~(\ref{eq:EBequation_real}) is negligible.
Then, the four terms in the R.H.S of Eq.~(\ref{eq:EBequation_real}) should be balanced. 
In addition, using Poisson's equation and Ampere's law in the gyrokinetic form\cite{Kato2024, Sugama1996, Sugama2009}, the sum of energy transfer over particle species is written as
\begin{equation}
\label{eq:sumQturb}
\sum_{a}{Q^{\rm turb}_a}
 =
\frac{1}{8\pi}\frac{\partial}{\partial t} \sum_{\bm{k}}\Bigg\langle\Bigg\langle \left( k^2+\lambda_D^{-2} \right)|\phi_{\bm{k}}|^2 - |\bm{B}_{\bm{k}}|^2 \Bigg\rangle\Bigg\rangle,\\
\end{equation}
where $\bm{B}_{\bm{k}}=i\bm{k}\times\bm{A}_{\bm{k}}$, and $\lambda_D=1/\sqrt{\sum_a{4\pi n_ae_a^2/T_a}}$ is the Debye length.
The R.H.S. of Eq.~(\ref{eq:sumQturb}) also becomes negligible under the transport time-scale ordering.
Therefore, the energy transfer from the turbulent fluctuation to particle species $a$, $Q^{\rm turb}_a=\sum_{\bm{k}}Q^{\rm turb}_{a\bm{k}}$, can be regarded as the energy exchange between different particle species.
\subsection{Energy flux and energy exchange driven by microturbulence}
\label{subsec:2B}
The turbulent energy transfer in the wavenumber space $Q^{\rm turb}_{a\bm{k}}$ can be divided into four components under the transport time-scale ordering, described as
\begin{eqnarray}
Q^{\rm turb}_{a\bm{k}}&=&Q^{\rm turb}_{a\parallel\bm{k}}+Q^{\rm turb}_{aB\bm{k}}+Q^{\rm turb}_{a\psi\bm{k}}+Q^{\rm turb}_{aC\bm{k}}, \label{eq:TEE_parts} \\
Q^{\rm turb}_{a\parallel\bm{k}}&=&\Re\Bigg\langle\Bigg\langle \int d^3v \bm{G}_{a\parallel\bm{k}}\cdot\bm{j}_{a\parallel\bm{k}}^* \Bigg\rangle\Bigg\rangle, \label{eq: parallel_heating}\\
Q^{\rm turb}_{aB\bm{k}}&=&\Re\Bigg\langle\Bigg\langle \int d^3v 
 \bm{G}_{a\perp\bm{k}}\cdot\bm{j}_{aB\bm{k}}^* \Bigg\rangle\Bigg\rangle,
 \label{eq:perp_heating}\\
Q^{\rm turb}_{a\psi\bm{k}}&=&\Re\Bigg\langle\Bigg\langle \int d^3v \bm{G}_{a\perp\bm{k}}\cdot \bm{j}_{a\psi\bm{k}}^* \Bigg\rangle\Bigg\rangle,
\label{eq:psi_heating}\\
Q^{\rm turb}_{aC\bm{k}}&=&-\Re\Bigg\langle\Bigg\langle \int d^3v C_a^{GK} \psi_{a\bm{k}}^* \Bigg\rangle\Bigg\rangle,
\label{eq:collision_psi}
\end{eqnarray}
where the perturbed fields $\bm{G}_{a\bm{k}}$ and
the perturbed currents $\bm{j}_{a\bm{k}}$ at the gyrocenter position are defined by
\begin{eqnarray}
\bm{G}_{a\bm{k}} 
& = & -\nabla_\parallel \psi_{a\bm{k}}-i\bm{k}\psi_{a\bm{k}}
=\bm{G}_{a\parallel\bm{k} }+\bm{G}_{a\perp\bm{k} } , \label{eq:Gfield}
\\ 
\bm{j}_{a\bm{k}} & = & 
\bm{j}_{a\parallel\bm{k} }+\bm{j}_{a\perp\bm{k} }, 
\\
\bm{j}_{a\parallel\bm{k}} 
& = &
e_ah_{a\bm{k}}v_\parallel\bm{b}, 
\\ \bm{j}_{a\perp\bm{k}}
& = & \bm{j}_{aB\bm{k}}+\bm{j}_{a\psi\bm{k}}, 
\\ 
\bm{j}_{aB\bm{k}} 
& = & 
e_ah_{a\bm{k}}\bm{v}_{da}, 
\\ 
\bm{j}_{a\psi\bm{k}}  &= & 
\frac{ice_a}{B} \sum_{\bm{k}'+\bm{k}''= 
\bm{k}}\left( \bm{b}\times\bm{k}'\right) \psi_{a\bm{k}'}h_{a\bm{k}''} .\label{eq:u_psi}
\end{eqnarray}
The collisional components $Q^{\rm turb}_{aC\bm{k}}(a=e, i)$ are negligible in weakly collisional plasmas, and the component of nonlinear interaction $Q^{\rm turb}_{a\psi\bm{k}}$ satisfies
\begin{equation}
\label{sumqpsi}
    \sum_{\bm{k}} Q^{\rm turb}_{a\psi\bm{k}}=0.
 \end{equation}
Then, the turbulent energy transfer $Q^{\rm turb}_a$ is determined by the Joule heating via currents resulting
from streaming motions parallel to the background magnetic field, $Q^{\rm turb}_{a \parallel}=\sum_{\bm{k}}Q^{\rm turb}_{a \parallel \bm{k}}$, and from $\nabla B$-curvature drifts perpendicular to it, $Q^{\rm turb}_{a B}=\sum_{\bm{k}}Q^{\rm turb}_{a B \bm{k}}$.
In low beta plasmas, the $\nabla B$-curvature drift can be expressed as 
 $\bm{v}_{da} =\bm{b}\times\left(v_\parallel^2+\mu B/m_a\right)\nabla B/\Omega_a B$. 
Then, 
under the electrostatic approximation,  $Q^{\mathrm{turb}}_{aB \bm{k}}$ in Eq.~(\ref{eq:perp_heating})  is rewritten as 
\begin{equation}
    Q^{\mathrm{turb}}_{aB \bm{k}} = 
\Re\Bigg\langle\Bigg\langle 2  P_{a\bm{k}}^* 
\left(
 \frac{c}{B} \bm{E}_{\bm{k}}\times \bm{b}\right)\cdot 
\nabla \ln{B}
\Bigg\rangle\Bigg\rangle
\approx -\frac{1}{R_0}\mathcal{E}^{\rm turb}_{a\bm{k}},
\label{eq:perp_heating_ITG}
\end{equation}
where $\bm{E}_{\bm{k}} \equiv -i \bm{k}  \phi_{\bm{k}}$ is the electric field perturbation and 
$P_{a \bm{k}} \equiv \int d^3v\left( m_a v_\parallel^2+\mu B\right) f_{a\bm{k}}/2 $ roughly represents the pressure perturbation. 
The scale length for the gradient of the magnitude of background magnetic field $|\left(\nabla\ln B\right)^{-1}|$ is approximated as the major radius $R_0$ here.
Accordingly, the perpendicular heating can be approximated as energy flux, defined as $\mathcal{E}^{\rm turb}_{a\bm{k}}=q^{\rm turb}_{a\bm{k}}+5T_a\Gamma^{\rm turb}_{a\bm{k}}/2$, divided by the major radius.
If we make a rough assumption that $Q^{\rm turb}_{e\parallel}\approx Q^{\rm turb}_{i\parallel}$ here, the difference in energy exchange between electrons and ions can be written as
\begin{eqnarray}
    Q^{\rm turb}_e-Q^{\rm turb}_{i}&=&\left(Q^{\rm turb}_{e\parallel}+Q^{\rm turb}_{eB}\right)-\left(Q^{\rm turb}_{i\parallel}+Q^{\rm turb}_{iB}\right) \nonumber \\
    &\approx& Q^{\rm turb}_{eB}-Q^{\rm turb}_{iB}\approx -\frac{\mathcal{E}^{\rm turb}_{e}-\mathcal{E}^{\rm turb}_{i}}{R_0},
\end{eqnarray}
where $\mathcal{E}^{\rm turb}_{a}=\sum_{\bm{k}}\mathcal{E}^{\rm turb}_{a\bm{k}}$.
Using the relation $Q^{\rm turb}_i=-Q^{\rm turb}_e$, the turbulent energy transfer for ions can be approximated as
\begin{equation}
    Q^{\rm turb}_i\approx\frac{\mathcal{E}^{\rm turb}_{e}-\mathcal{E}^{\rm turb}_{i}}{2R_0}.
    \label{eq:Qturb_EFlux}
\end{equation}
From Eqs.~(\ref{eq:perp_heating_ITG}) and (\ref{eq:Qturb_EFlux}), the parallel heating for ions is also approximated as
\begin{equation}
    Q^{\rm turb}_{i \parallel}\approx\frac{\mathcal{E}^{\rm turb}_{e}+\mathcal{E}^{\rm turb}_{i}}{2R_0}.
    \label{eq:Qpara_Eflux}
\end{equation}
Whether these approximations are valid in ITG-TEM turbulence is investigated in Sec.~\ref{subsec:3C}.
\subsection{Quasilinear model}
\label{subsec:2D}
The quasilinear model estimates the nonlinear simulation results based on two components:
(i) the ratio of fluxes or energy exchange to the squared amplitude of the electrostatic potential, obtained from linear simulations, and
(ii) the saturation rule of the squared amplitude of the electrostatic potential in a steady state turbulence.
In this study, we focus on the former and do not deal with the latter.
The ratio is calculated by
\begin{eqnarray}
WX_{a,N}(k_y)&=&C_X\frac{\sum_{k_x}\langle \langle \tilde{X}_{a\bm{k},N}\rangle\rangle}{\sum_{k_x}\langle \langle |\phi_{\bm{k}, N}|^2 \rangle\rangle}, \label{eq:WX_N}\\
WX^{0}_{a,N}(k_y)&=&C_X\frac{\langle \langle \tilde{X}_{ak_x=0, k_y,N}\rangle\rangle}{\langle \langle |\phi_{k_x=0, k_y, N}|^2 \rangle\rangle},  \label{eq:WX^0_N}\\
WX^{0}_{a,L}(k_y)&=&C_X\frac{\langle \tilde{X}_{ak_x=0, k_y,L}\rangle_s}{\langle |\phi_{k_x=0, k_y, L}|^2 \rangle_s}, \label{eq:WX^0_L}
\end{eqnarray}
where $X_a$ indicates $(\Gamma_a, q_a, Y_a)$.
The superscript $0$ and subscript $L, N$ denote the results obtained using only the $k_x=0$ modes 
 (instead of summing over the $k_x$ space), those from linear and nonlinear calculations, respectively.
Normalization factors $C_X$, and the real parts of integrals inside the double average over the ensemble and the flux surface $\langle \langle \cdots \rangle \rangle$ of Eqs.~(\ref{eq:Particle Heat fluxes}), $\tilde{X}_{a\bm{k}}=(\tilde{\Gamma}_{a\bm{k}}, \tilde{q}_{a\bm{k}}, \tilde{Y}_{a\bm{k}})$, are described as 
\begin{eqnarray}
    & &\left(C_\Gamma, C_q, C_Y\right)=\left(\frac{T_i^2}{e^2n_ev_{ti}}, \frac{T_i}{e^2n_ev_{ti}}, \frac{T_iR_0}{e^2n_ev_{ti}}\right), \\
    & &\tilde{\Gamma}_{a\bm{k}}=\Re\left[  \int d^3v h_{a\bm{k}}^*\left(-i\frac{c}{B}\psi_{a\bm{k}}\bm{k}\times\bm{b}\right)\cdot\nabla r \right], \\
    & &\tilde{q}_{a\bm{k}}=\Re\left[  \int d^3v \left(w-\frac{5T_a}{2}\right) h_{a\bm{k}}^*\left(-i\frac{c}{B}\psi_{a\bm{k}}\bm{k}\times\bm{b}\right)\cdot\nabla r \right], \nonumber\\
    \\
    & &\tilde{Y}_{a\bm{k}}=\Re\left[\frac{e_a}{2} \int d^3v \left(h_{a\bm{k}}^*\frac{\partial\psi_{a\bm{k}}}{\partial t} - \frac{\partial h^*_{a\bm{k}}}{\partial t} \psi_{a\bm{k}}\right)\right], 
\end{eqnarray}
respectively. 
The expression for turbulent energy transfer is reformulated as $Y_{a\bm{k}}$, as developed in Refs.~\cite{Candy, Kato2024} .
This formulation preserves the condition of energy exchange $\sum_aY_{a\bm{k}}=0$ even in linear simulations, which is suitable for the quasilinear model, as reported in Ref.~\cite{Kato2024}.
In this study, we call $WX_{a,N}(k_y)$ and $WX^{0}_{a,L}(k_y)$, nonlinear weights and quasilinear weights, respectively.

\begin{table*}[t]
\centering
    \caption{Plasma and field parameters}
\begin{tabular}{lcc} \hline
     \multicolumn{2}{c}{Plasma and field parameters}&   Value \\ \hline
    Normalized electron and ion temperature gradients & $(R_0/L_{Te}, R_0/L_{Ti})$  &  
    \begin{tabular}{l}
    (7.0, 1.0), (6.0, 2.0), (5.0, 3.0), (4.0, 4.0),\\
    (3.0, 5.0), (2.0, 6.0), (1.0, 7.0) 
    \end{tabular} \\
    Normalized density gradient& $R_0/L_{na}$  &   2.22    \\
    Mass ratio & $m_e/m_i$  &   $5.446\times10^{-4}$    \\
    Charge ratio& $e_e/e_i$  &   -1    \\
    Ion beta value& $\beta_i= 4 \pi n_iT_i/B_{ax}^2$  &   $1\times10^{-4}$    \\
    Collision frequency for ions& $\nu^*_{ii}\equiv R_0 q_0 \nu_{ii}/(\epsilon_r^{3/2}v_{ti})$  &   0.068    \\
    Collision frequency for electrons& $\nu^*_{ee}\equiv R_0 q_0 \nu_{ee}/(\epsilon_r^{3/2}v_{te})$  &   0.007\\
    Temperature ratio& $T_e/T_i$  &   $3.0$    \\ 
    Inverse aspect ratio& $\varepsilon_r$  &   $0.18$    \\ 
    Safety factor & $q_0$ & $1.4$ \\
    Magnetic shear & $\hat{s}=(r/q)(dq/dr)$ & $0.78$ \\
    \hline
    \label{tab:Plasma and field parameters}
\end{tabular}
\end{table*}

\begin{table}
    \caption{Resolution settings}
\begin{tabular}{c} \hline
     Domain sizes and \\ resolved perpendicular wavenumbers \\ \hline
    $-64.10\rho_{ti}\leq x \leq 64.10\rho_{ti}$\\
    $-62.83\rho_{ti}\leq y \leq 62.83\rho_{ti}$   \\
    $-\pi \leq z\leq \pi$    \\
    $-4v_{ta}\leq v_\parallel \leq 4v_{ta}$    \\
    $0\leq \mu B_0/T_a \equiv m_av_\perp^2/2T_a\leq 8$    \\ 
$-4.70\rho_{ti}^{-1}\leq k_x\leq4.70\rho_{ti}^{-1}, (k_{x,\mathrm{min}}=0.049\rho_{ti}^{-1})$ \\
$-1.55\rho_{ti}^{-1}\leq k_y\leq1.55\rho_{ti}^{-1}, (k_{y,\mathrm{min}}=0.050\rho_{ti}^{-1})$ \\ \hline
\\ 
\hline
Grid points in $(x, y, z, v_\parallel, \mu)$\\ \hline
$288\times96\times96\times64\times32$ \\ \hline
\label{tab:resolution}
\end{tabular}
\end{table}

\section{Simulation Results}\label{sec:3}
\subsection{Simulation settings}\label{subsec:3A}

\begin{figure*}[tb]
    \centering
    \includegraphics[keepaspectratio, scale=0.45]{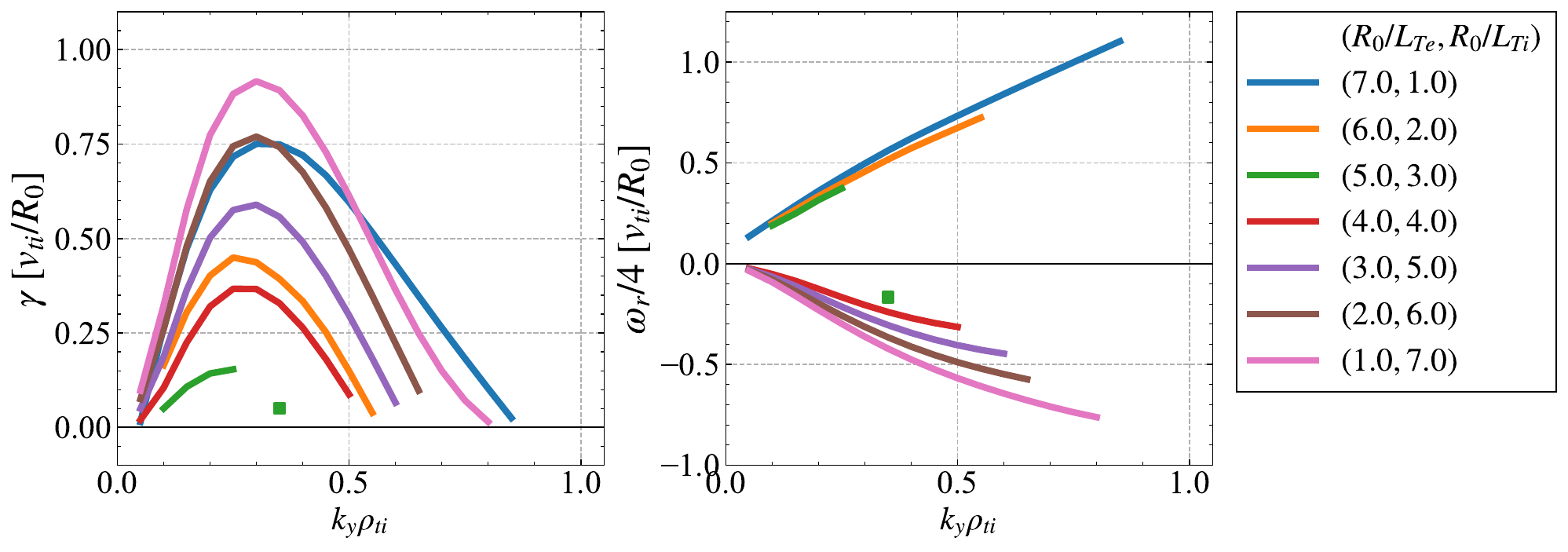}
    \caption{The linear growthrate $\gamma$ (left) and frequency $\omega_r$ (right) at $k_x\rho_{ti}=0$.
    %
    %
    The most unstable mode transitions between ITG and TEM around $(R_0/L_{Te}, R_0/L_{Ti})=(5.0, 3.0)$.}
    \label{fig:LinearSimulation}
\end{figure*}

In this study, microturbulence simulations are conducted using the GKV code \cite{GKV}, which numerically solves the gyrokinetic equations within the flux-tube domain. 
Table \ref{tab:Plasma and field parameters} shows plasma and field parameters, and Tab. \ref{tab:resolution} shows resolution settings used in this work. 
We here investigate ITG-TEM turbulence, and set the temperature ratio $T_e/T_i=3.0$ to suppress the ETG instability~\cite{Nakata}.
The normalized electron and ion temperature gradients, $R_0/L_{Te}$ and $R_0/L_{Ti}$, are varied in order to modify the dominant instability while keeping the total pressure gradient, $\sum_{a=e, i}(R_0/L_{na}+R_0/L_{Ta})$, constant.
The pressure-gradient component of the drift velocity is not neglected but remains constant in this study.
To focus on pure TEM turbulence and suppress instabilities in high wavenumber modes, we set $(R_0/L_{Te}, R_0/L_{Ti})=(7.0, 1.0)$ instead of using $R_0/L_{Ti}=0.0$.
We set the normalized density gradients $R_0/L_{na}=2.22(a=e, i)$, which is the same parameter used in the CBC case\cite{Dimits}.
The flux tube coordinates used here are $x=r-r_0, y=r_0/q_0\left(q(r)\theta-\zeta\right)$, and $z=\theta$ in low-$\beta$, large aspect ratio, axisymmetric tokamak plasmas with circular, concentric flux surfaces, where $r$, $\theta$, and $\zeta$ denote the minor radius, poloidal angle, and toroidal angle, respectively.
The subscript $0$ signifies the parameters in the center of the flux tube.

The perpendicular wavenumber is expressed as $\bm{k}=\left( k_x +\hat{s} z k_y\right) \bm{e}_r +k_y \bm{e}_\theta$. Here, $k_x$ and $k_y$ represent wavenumbers along the directions of $\nabla x$ and $\nabla y$, respectively, while $\bm{e}_r$ and $\bm{e}_\theta$ are unit vectors aligned with $\nabla r$ and $\nabla \theta$, respectively.
The ion beta value is set to $\beta_i= 4 \pi n_iT_i/B_{ax}^2=1\times 10^{-4}$, ensuring the validity of the electrostatic approximation.
Here, $B_{ax}$ denotes the magnetic field strength at the magnetic axis.
In these simulations, $B_{\parallel \bm{k}}$ is neglected, whereas $A_{\parallel \bm{k}}$ is retained to prevent numerical difficulties arising from very rapid electrostatic waves known as the $\omega_H$ mode~\cite{Lee}.
The Lenard-Bernstein collision operator~\cite{Lenard} is employed due to its computational efficiency compared to more sophisticated collision models.
Nevertheless, we anticipate that the choice of collision model does not significantly affect the results of this study, given that the normalized collision frequency is set to $\nu^*_{ii}\equiv R_0 q \nu_{ii}/(\epsilon_r^{3/2}v_{ti}) = 0.068 \ (R_0\nu_{ii}/v_{ti}=3.7\times10^{-3})$, which is considerably smaller than the growth rates of the instabilities examined here.
The thermal velocity for particle species $a$ is defined as $v_{ta}\equiv \sqrt{T_a/m_a}$.
Figure~\ref{fig:LinearSimulation} shows linear growth rates and real frequencies obtained by linear simulations varying the temperature gradients.
The sign of real frequencies reverses around $(R_0/L_{Te}, R_0/L_{Ti})=(5.0, 3.0)$, indicating that the most unstable mode transitions between ITG and TEM.
For the case of $(R_0/L_{Te}, R_0/L_{Ti}) = (5.0, 3.0)$, the most unstable mode transitions between ITG and TEM around $k_y\rho_{ti} = 0.30$.
In this region, ITG and TEM modes are comparably unstable, and therefore the solution of the initial-value problem does not converge to a single eigenmode of the form $\exp[(-i\omega_r+\gamma) t]$.
The $k_y\rho_{ti} = 0.30$ mode corresponds to this transition boundary and is not plotted in Fig.~\ref{fig:LinearSimulation}.
Nonlinear simulations are performed using these plasma parameters to investigate the direction, magnitude, underlying physical mechanisms, and predictability of quasilinear models for turbulent energy exchange in ITG-TEM turbulence.
Although the traditional twist-and-shift flux-tube model is known to affect the nonlinear simulation results due to the self-interaction \cite{Ball}, this issue does not significantly alter our analysis and conclusions.

\subsection{Turbulent energy exchange in pure TEM turbulence}\label{subsec:3B}

\begin{figure}[bp]
    \includegraphics[keepaspectratio, scale=0.55]{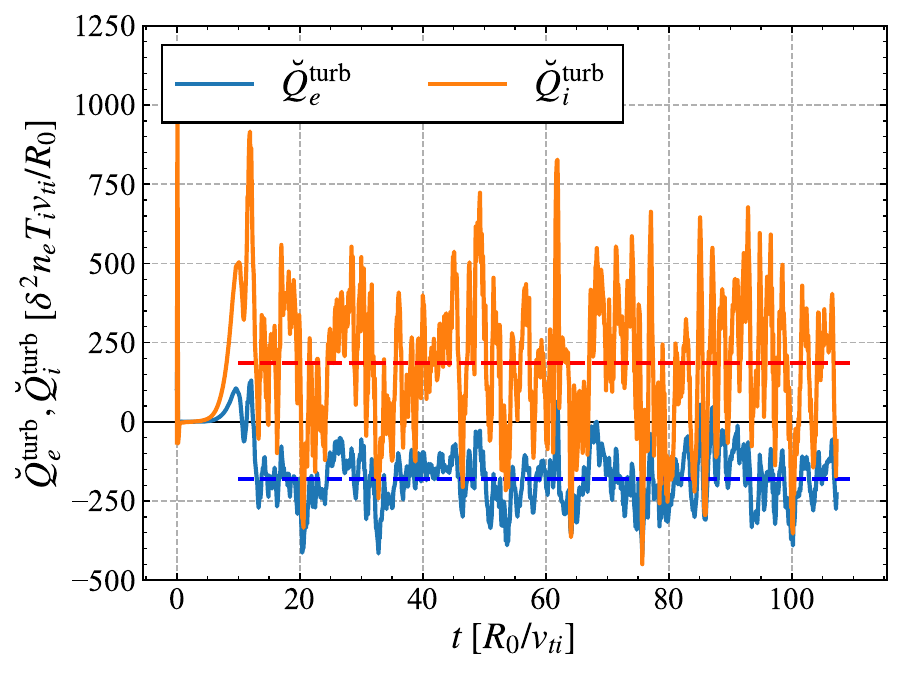}
    \caption{Time evolution of instantaneous turbulent energy transfers $\breve{Q}^{\rm turb}_{a} (a=e, i)$ in pure TEM turbulence. Red and blue dashed lines indicate energy transfer averaged over time in a steady state turbulence for ions and electrons, $Q^{\rm turb}_{i}$ and $Q^{\rm turb}_{e}$, respectively.}
    \label{fig:EnergyexchangepureTEM}
\end{figure}

\begin{figure}[tbp]
    \includegraphics[keepaspectratio, scale=0.42]{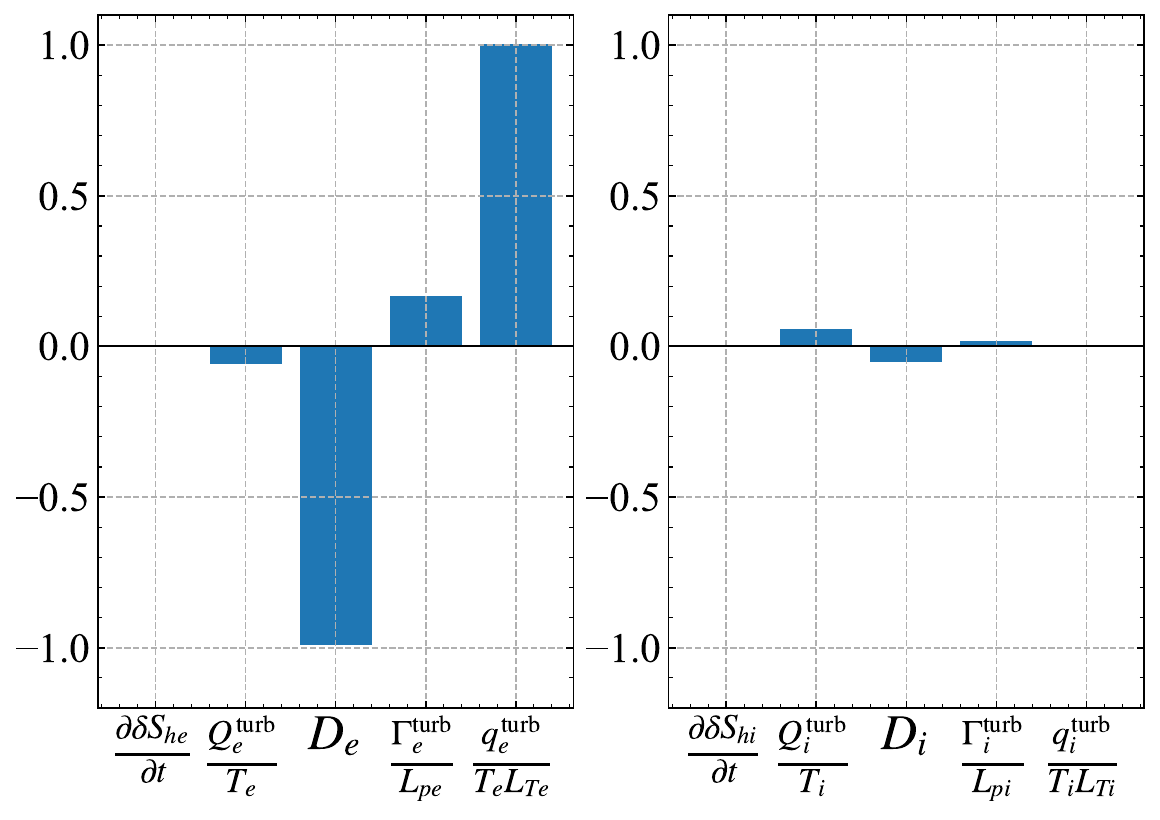}
    \caption{Comparison of all terms in the entropy balance equation, Eq.~(\ref{eq:EBequation_real}), in a steady state of the pure TEM turbulence for $(R_0/L_{Te}, R_0/L_{Ti})=(7.0, 1.0)$. 
All terms in Eq.~(\ref{eq:EBequation_real}) for electrons (left) and ions (right) are normalized by $q_e^{\rm turb}/(T_eL_{Te})$ and $q_e^{\rm turb}/(T_iL_{Te})$, respectively.
}\label{fig:EntropyBalance_pureTEM}
\end{figure}

\begin{figure*}[tbp]
    \includegraphics[keepaspectratio, scale=0.50]{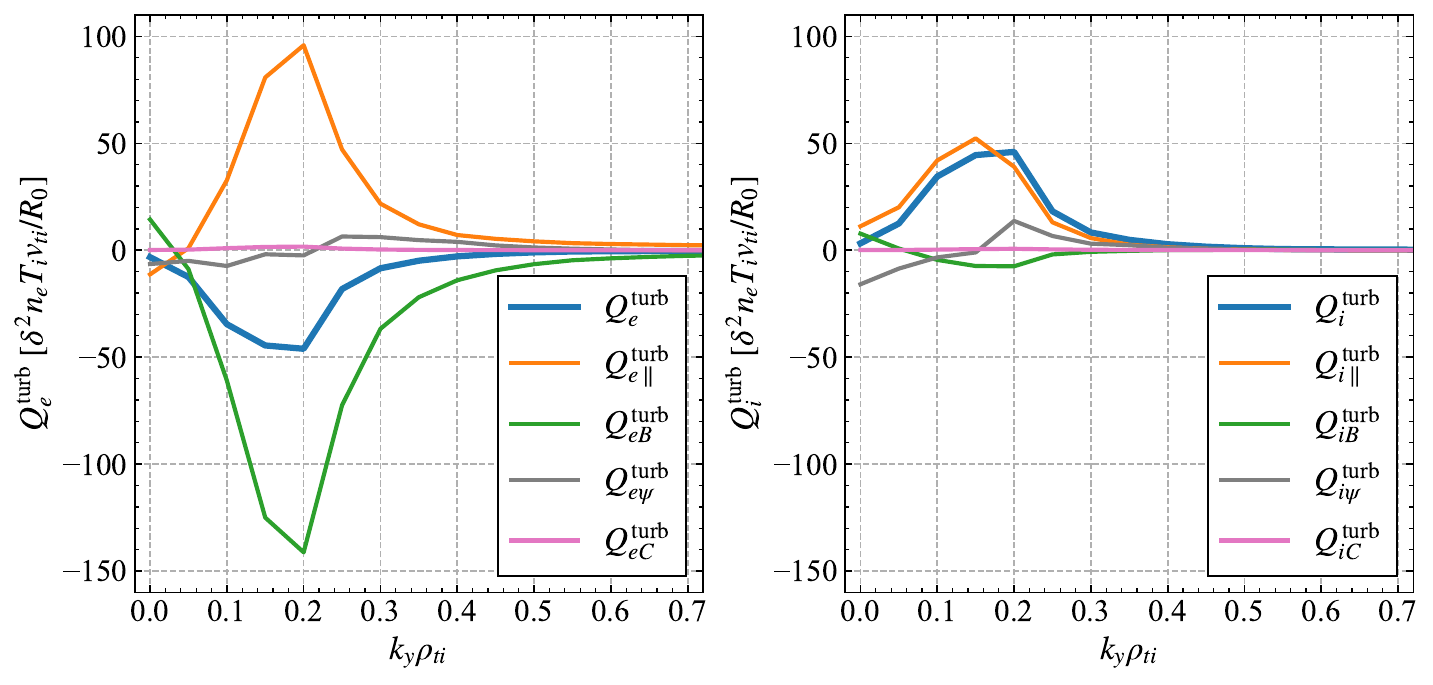}
    \caption{The wavenumber spectra of turbulent energy transfer terms in Eqs.~(\ref{eq: parallel_heating})--(\ref{eq:collision_psi}) for electrons (left) and ions (right) in the case of $(R_0/L_{Te}, R_0/L_{Ti}) =(7.0, 1.0)$. 
The spectra are given as functions of $k_y \rho_{ti}$ obtained by summing over $k_x$. 
The electron cooling due to the $\nabla B$-curvature drift denoted by $Q_{eB}^{\rm turb}<0$ and the ion heating due to the parallel field denoted by  $Q_{i\parallel}^{\rm turb}>0$ are dominant mechanisms in the turbulent energy exchange between electrons and ions in TEM turbulence. 
}
    \label{fig:partQ_pureTEM}
\end{figure*}

The nonlinear simulation results for pure TEM turbulence with $(R_0/L_{Te}, R_0/L_{Ti})=(7.0, 1.0)$ are presented here.
The instantaneous energy transfer is given by 
\begin{equation}
    \breve{Q}^{\rm turb}_{a}=\sum_{\bm{k}}\Re\Bigg\langle \int d^3v e_ah_{a\bm{k}}^*\frac{\partial\psi_{a\bm{k}}}{\partial t} \Bigg\rangle_s, 
\end{equation}
where $\langle \cdots\rangle_s$ denotes the surface average without ensemble or time averaging.
The time evolution of instantaneous energy transfer $\breve{Q}^{\rm turb}_{a}$ in pure TEM turbulence is shown in Fig.~\ref{fig:EnergyexchangepureTEM}.
The instantaneous energy transfer associated with ions $\breve{Q}^{\rm turb}_{i}$ exhibits larger spike-like structures compared to that of electrons $\breve{Q}^{\rm turb}_{e}$, and the instantaneous net turbulent heating $\sum_{a=e, i} \breve{Q}^{\rm turb}_{a}$ is not necessarily zero at every moment in time.
Over a longer timescale in a steady state, however, the sum approaches zero, indicating that there is no net energy generation and that energy is being exchanged between electrons and ions.
The time average in a steady state of turbulence is used instead of the ensemble average to evaluate $\langle\langle\cdots\rangle\rangle$.
In the steady state, the time-averaged turbulent heatings $Q^{\rm turb}_e$ and $Q^{\rm turb}_i$ are $-180$ and $184~[\delta^2n_eT_iv_{ti}/R_0]$, respectively, with standard deviations of $84$ and $210~[\delta^2n_eT_iv_{ti}/R_0]$.
Thus, in the magnetic configuration of CBC, energy is transferred from electrons to ions in TEM turbulence, which is in the direction opposite to that in ITG turbulence.
Therefore, TEM turbulence in tokamak palasmas is expected to facilitate energy transfer from alpha-heated electrons to ions, potentially enhancing the efficiency of ion heating.
Figure~\ref{fig:EntropyBalance_pureTEM} shows all terms in Eq.~(\ref{eq:EBequation_real}) in a steady state of turbulence.
For clarity, each term for electrons and ions is normalized by $q^{\mathrm{turb}}_e/(T_eL_{Te})$ and $q^{\mathrm{turb}}_e/(T_iL_{Te})$, respectively.
The numerical error in the entropy balance, defined as the difference between the left- and right-hand sides of Eq.~(\ref{eq:EBequation_real}) is within 12 \% of $q^{\mathrm{turb}}_e/(T_eL_{Te})$ for electrons and 3 \% for ions, while $Q^{\mathrm{turb}}_i=-Q^{\mathrm{turb}}_e$ is 6 \%.
This error arises from the numerical diffusion as well as from losses due to outflow at the $z$-coordinate boundaries.
When the resolution in the $z$-coordinate is doubled, the error associated with the numerical diffusion is reduced by 1/16.
Although these issues can be mitigated by increasing the resolution in the $z$-coordinate and extending the $z$-domain, the resolution and domain employed in this study are sufficient for evaluating fluxes and energy exchanges\cite{Kato2024}.
For electrons, particle and heat fluxes generate entropy, whereas energy exchange and collisions reduce entropy for turbulent fluctuations $\delta S_{he}$, thereby maintaining balance.
In contrast, for ions, energy exchange significantly contributes to entropy production in addition to the ion particle flux.
The entropy of turbulent fluctuations in the ion distribution function is kept in a steady state due to the collisional dissipation.

This situation is analogous to that in ITG turbulence in terms of transferring energy from particle species with large entropy production due to turbulent transport to particle species with small entropy production.
In TEM turbulence, it is found that electrons generate larger entropy through turbulent transport than ions, and the direction of energy exchange is from electrons to ions.
Thus, it is confirmed that the conjecture regarding the direction of turbulent energy transfer, as discussed in Sec.~\ref{sec:2}, is applicable to pure TEM turbulence as well.
The wavenumber spectrum of each component of turbulent energy transfer in Eq.~(\ref{eq:TEE_parts}) is shown in Fig.~\ref{fig:partQ_pureTEM}.
First, the collisional components $Q^{\mathrm{turb}}_{aC} (a=e, i)$ are found to have negligible influence on the turbulent energy exchange.
Similarly, the nonlinear interaction components between different wavenumbers, $Q^{\mathrm{turb}}_{a\psi}(a=e, i)$, have small impact across the wavenumber spectrum.
As in ITG turbulence, both parallel and perpendicular heating terms, $Q^{\mathrm{turb}}_{a\parallel}$ and $Q^{\mathrm{turb}}_{aB}(a=e, i)$, predominantly determine the magnitude and direction of energy transfer in TEM turbulence.
However, the magnitudes of these contributions for electrons and ions differ from those in ITG turbulence.
The turbulent energy transfer in ITG turbulence is mainly composed of perpendicular cooling for ions $Q^{\mathrm{turb}}_{iB}<0$ and parallel heating for electrons $Q^{\mathrm{turb}}_{e\parallel}>0$.
In contrast, the energy transfer in TEM turbulence is dominated by perpendicular cooling for electrons $Q^{\mathrm{turb}}_{eB}<0$ and parallel heating for ions $Q^{\mathrm{turb}}_{i\parallel}>0$, as also clearly shown in Fig.~\ref{fig:partQ_LTi_mixedITGTEM}.
These results suggest that the signs of parallel heating $Q^{\rm turb}_{a\parallel \bm{k}}>0\ (a=e, i)$ and perpendicular cooling $Q^{\rm turb}_{aB \bm{k}}<0\ (a=e, i)$ except for zonal mode ($k_y=0$) remain unchanged when the dominant instability changes.
This indicates that the direction of turbulent energy exchange is determined by the relative magnitudes of these components.

\subsection{Turbulent energy exchange in mixed ITG-TEM turbulence}\label{subsec:3C}

\begin{figure*}[tbp]
    \includegraphics[keepaspectratio, scale=0.39]{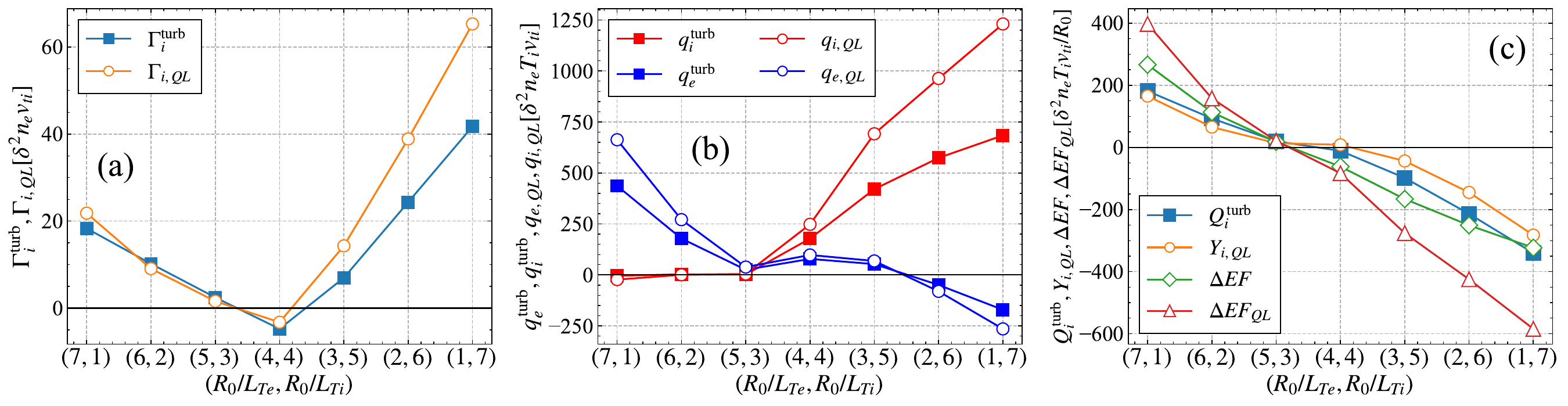}
    \caption{The turbulent particle flux $\Gamma^{\mathrm{turb}}_i(=\Gamma^{\mathrm{turb}}_e)$, electron and ion heat fluxes, $q^{\mathrm{turb}}_e$ and $q^{\mathrm{turb}}_i$, and energy transfer $Q^{\mathrm{turb}}_i(=-Q^{\mathrm{turb}}_e)$ as functions of $(R_0/L_{Te}, R_0/L_{Ti})$.
    Square and circular markers in Figs.~\ref{fig:PF_HF_EE_LTi_ITGTEM}~(a-c) indicate nonlinear results and results calculated by Eq.~(\ref{eq:QuasilinearFluxes}), respectively.
    Diamond and triangle markers in Fig.~\ref{fig:PF_HF_EE_LTi_ITGTEM}~(c) represent the results of Eq.~(\ref{eq: delta EF}) and Eq.~(\ref{eq:DeltaEF}), respectively.
    }
    \label{fig:PF_HF_EE_LTi_ITGTEM}
\end{figure*}

\begin{figure*}[tbp]
    \includegraphics[keepaspectratio, scale=0.4]{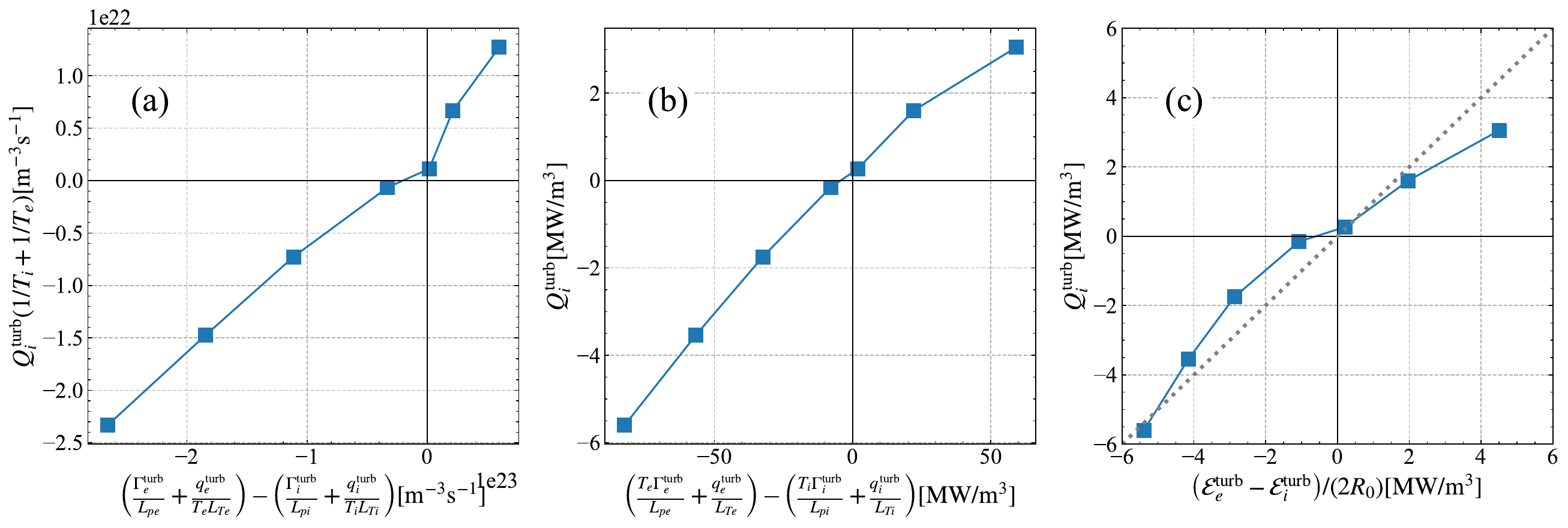}
    \caption{
    Turbulent energy transfer in mixed ITG-TEM turbulence as a function of $\Delta EP$ (a), $\Delta FP$ (b), and $\Delta EF$ (c), as defined in Eqs.~(\ref{eq: delta EP})-(\ref{eq: delta EF}), respectively.
    The physical dimensional quantities are calculated by using $\delta\equiv \rho_{ti}/R_0= 1.4\times10^{-3},$ $ n_e=1.0\times10^{20}\ \mathrm{m^{-3}},$ $ T_i=2.0\ \mathrm{keV}$.
    The dot line in Fig.~\ref{fig:Q_Entropy}(c) represents $Q^{\rm turb}_i=(\mathcal{E}^{\rm turb}_e-\mathcal{E}^{\rm turb}_i)/(2R_0)$.
}
    \label{fig:Q_Entropy}
\end{figure*}

\begin{figure*}[t]
    \includegraphics[keepaspectratio, scale=0.42]{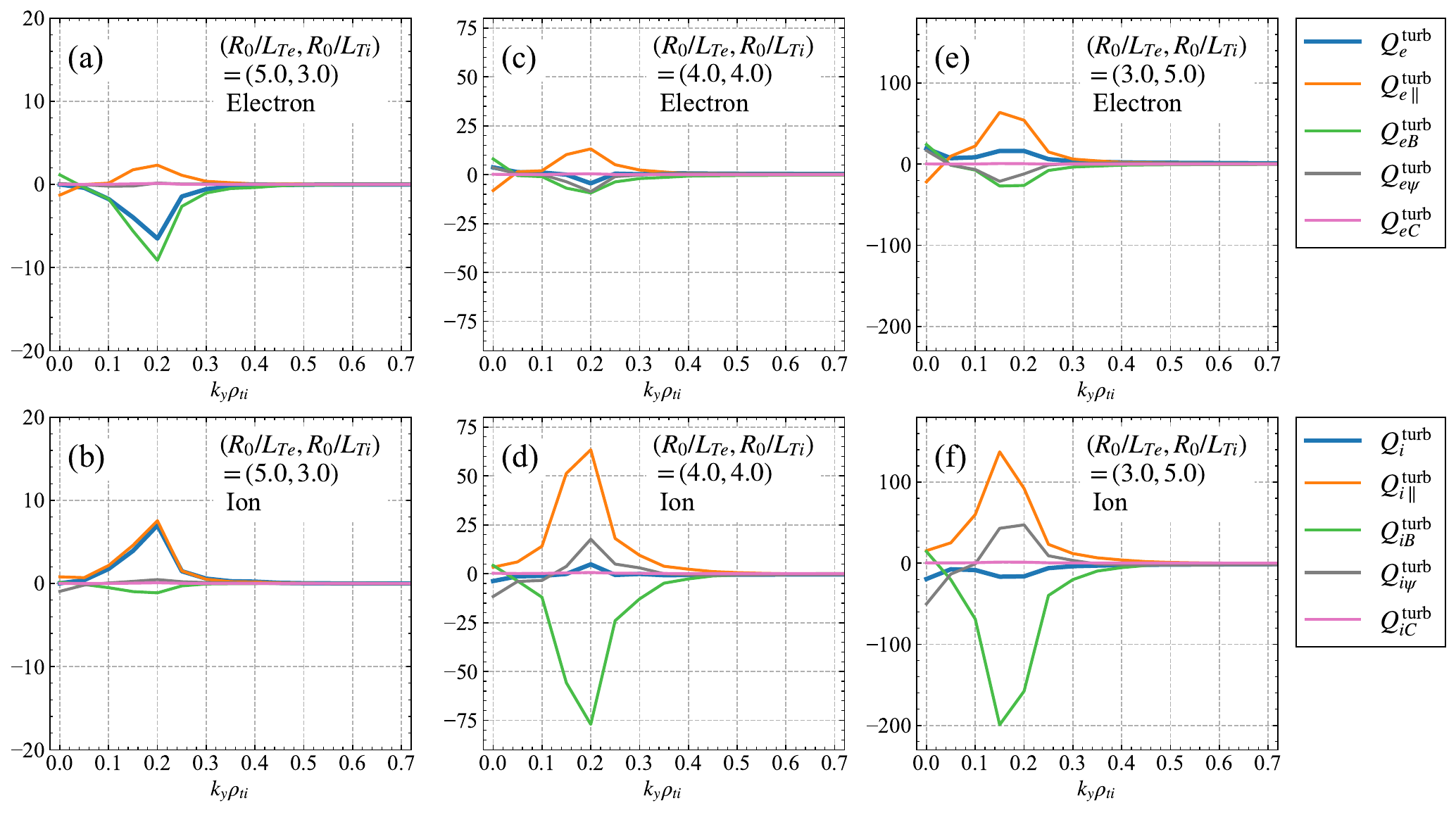}
    \caption{
    The wavenumber spectra of turbulent energy transfer terms in Eqs.~(\ref{eq: parallel_heating})--(\ref{eq:collision_psi}) for electrons (upper) and ions (lower) in the case of $(R_0/L_{Te}, R_0/L_{Ti})=(5.0, 3.0)$, $(4.0, 4.0)$, and $(3.0, 5.0)$. 
}
    \label{fig:partQspectrum_mixedITGTEM}
\end{figure*}

\begin{figure}[tbp]
    \includegraphics[keepaspectratio, scale=0.26]{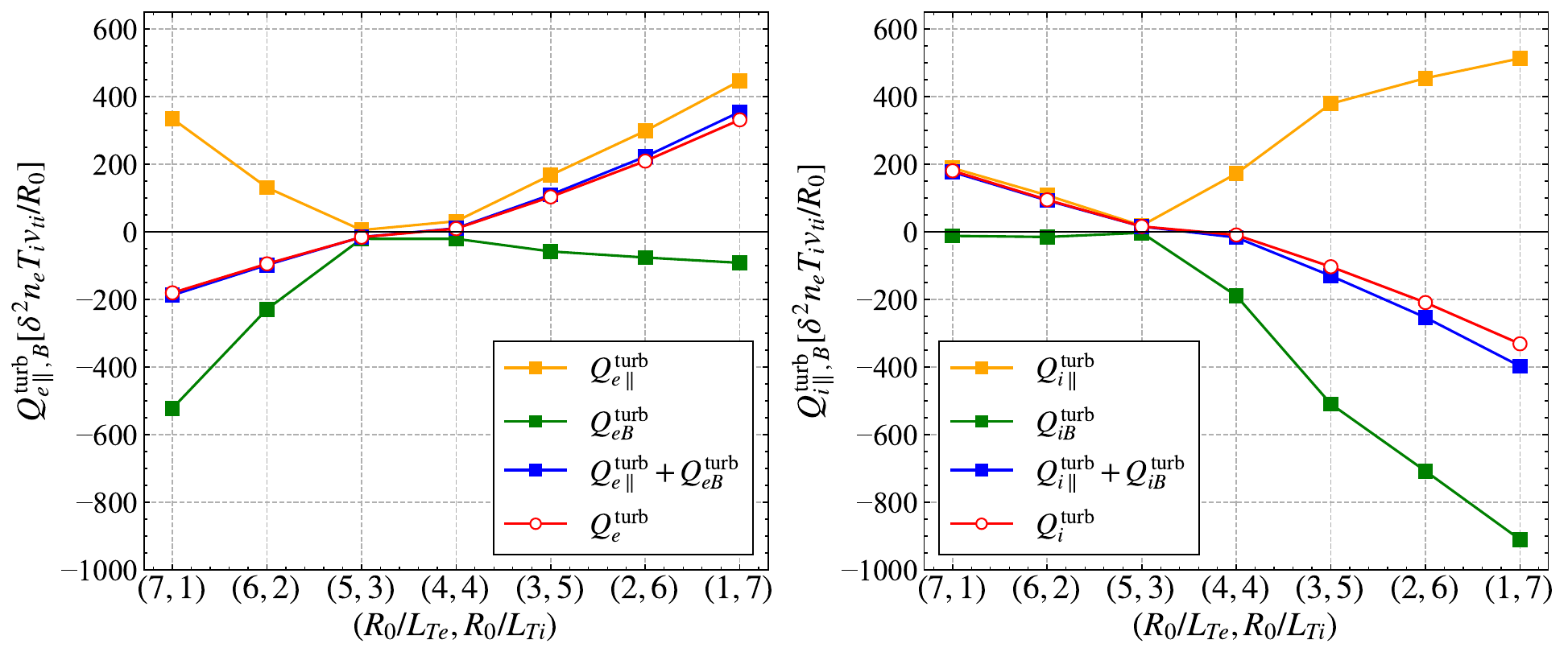}
    \caption{
The parallel heating and perpendicular cooling, $Q^{\rm turb}_{a \parallel}$ and $Q^{\rm turb}_{a B}$, for electrons (left) and ions (right) as functions of $(R_0/L_{Te}, R_0/L_{Ti})$.
The sum of the parallel heating and perpendicular cooling, as well as the total turbulent energy transfers, are also shown.
}
    \label{fig:partQ_LTi_mixedITGTEM}
\end{figure}

\begin{figure}[tbp]
    \includegraphics[keepaspectratio, scale=0.32]{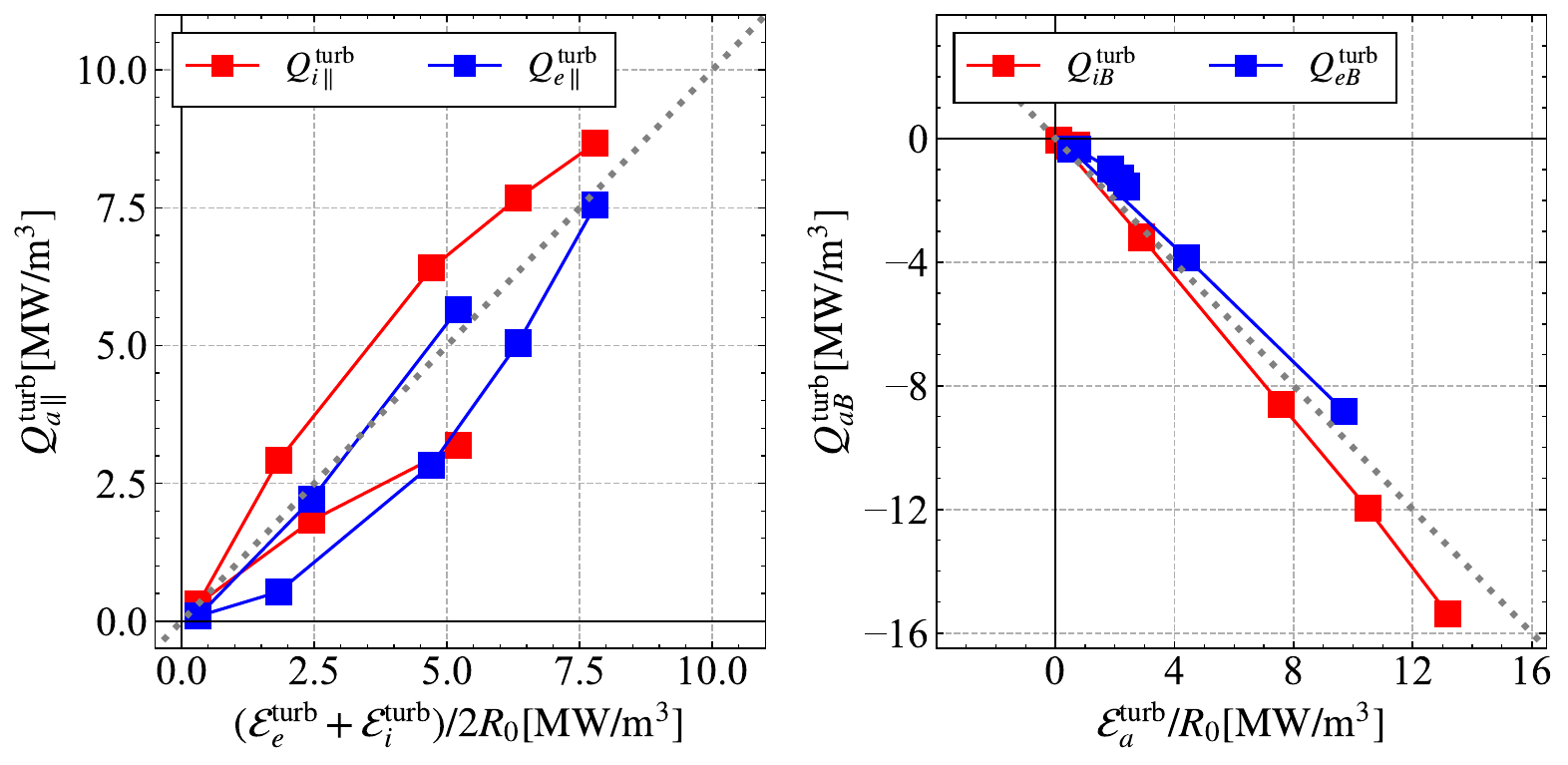}
    \caption{
For electrons and ions $(a=e, i)$, the parallel heating $Q^{\rm turb}_{a \parallel}$ (left) and perpendicular cooling $Q^{\rm turb}_{a B}$ (right) are plotted as functions of  $(\mathcal{E}^{\mathrm{turb}}_e+\mathcal{E}^{\mathrm{turb}}_i)/2R_0$, and $\mathcal{E}^{\mathrm{turb}}_a/R_0$, respectively. The dot lines represent $Q^{\rm turb}_{a\parallel}=(\mathcal{E}^{\rm turb}_e+\mathcal{E}^{\rm turb}_i)/(2R_0)$ and $Q^{\rm turb}_{aB}=-\mathcal{E}^{\rm turb}_a/R_0$, respectively.
}
    \label{fig:partQ_EF_mixedITGTEM}
\end{figure}

Next, we examine mixed ITG-TEM turbulence, where both ion and electron modes exist simultaneously.
Figure~\ref{fig:PF_HF_EE_LTi_ITGTEM} shows the turbulent particle flux $\Gamma^{\mathrm{turb}}_i(=\Gamma^{\mathrm{turb}}_e)$, electron and ion heat fluxes, $q^{\mathrm{turb}}_e$ and $q^{\mathrm{turb}}_i$, and energy transfer from electrons to ions $Q^{\mathrm{turb}}_i(=-Q^{\mathrm{turb}}_e)$ as functions of $(R_0/L_{Te}, R_0/L_{Ti})$.
It can be seen from Fig.~\ref{fig:PF_HF_EE_LTi_ITGTEM} that the particle flux, heat fluxes, and the absolute value of energy exchange become small around $(R_0/L_{Te}, R_0/L_{Ti}) = (5.0, 3.0)$ and $(4.0, 4.0)$, where the linear growth rate is lower than in other conditions.
As the instability mode approaches either the pure ITG or pure TEM regime, the particle flux and the absolute value of energy exchange increase.
When the electron temperature gradient $R_0/L_{Te}\ge5.0$, the electron heat flux increases, whereas the ion heat flux increases when the ion temperature gradient $R_0/L_{Ti}\ge4.0$.
The sign of the energy transfer for ions $Q^{\rm turb}_i$ is negative in ITG-dominated turbulence and positive in TEM-dominated turbulence.
In contrast to particle and heat fluxes, the energy exchange includes a contribution from the zonal mode ($k_y=0$).
While the zonal mode contribution is small in TEM-dominated turbulence, it accounts for approximately 10\% of the energy transfer in ITG-mode turbulence.
Figure~\ref{fig:Q_Entropy}(a) shows the turbulent energy transfer as a function of the difference in entropy production between electrons and ions $\Delta EP$, defined as
\begin{equation}
    \Delta EP\equiv\left(\frac{\Gamma^{\rm turb}_e}{L_{pe}}+\frac{q^{\rm turb}_{e}}{T_eL_{Te}}\right)-\left(\frac{\Gamma^{\rm turb}_i}{L_{pi}}+\frac{q^{\rm turb}_{i}}{T_iL_{Ti}}\right).
\label{eq: delta EP}
\end{equation}
In order to examine the conjecture about the relationship between the direction of energy exchange and entropy balance, it is necessary to identify which particle species produces larger entropy.
Therefore, $\Delta EP$ is set to the parameters for the horizontal axis.
In addition, the difference in the product of temperature and entropy production between electrons and ions $\Delta FP$, defined as
\begin{equation}
    \Delta FP\equiv \left(\frac{T_e\Gamma^{\rm turb}_e}{L_{pe}}+\frac{q^{\rm turb}_{e}}{L_{Te}}\right)-\left(\frac{T_i\Gamma^{\rm turb}_i}{L_{pi}}+\frac{q^{\rm turb}_{i}}{L_{Ti}}\right),
\label{eq: delta FP}
\end{equation}
as well as half the difference in energy flux between electrons and ions divided by the major radius $\Delta EF$, defined as
\begin{equation}
    \Delta EF\equiv\frac{\mathcal{E}^{\rm turb}_e-\mathcal{E}^{\rm turb}_i}{2R_0},
\label{eq: delta EF}
\end{equation}
are also used on the horizontal axis for comparison in Figs.~\ref{fig:Q_Entropy}~(b) and (c), respectively.
As illustrated in Fig.~\ref{fig:Q_Entropy}~(a), the sign of energy exchange is consistent with that of $\Delta EP$.
Therefore, this investigation indicates that energy is transferred from a particle species with larger entropy production due to turbulent transport to the other species in mixed ITG-TEM turbulence, and the validity of the conjecture is confirmed.
Furthermore, the signs of $\Delta FP$ and $\Delta EF$ are also found to coincide with that of the energy exchange, suggesting a linear correlation between energy exchange and them.
Since the slope of energy transfer as a function of $\Delta EF$ is approximately unity, this result supports the validity of Eq.~(\ref{eq:Qturb_EFlux}).
We next investigate the components of energy transfer in ITG-TEM turbulence.
Figures~\ref{fig:partQspectrum_mixedITGTEM} (a-f) show the wavenumber spectra of each component of turbulent energy transfer in Eqs.~(\ref{eq: parallel_heating})--(\ref{eq:collision_psi}) for electrons and ions in the case of $(R_0/L_{Te}, R_0/L_{Ti})=(5.0, 3.0)$, $(4.0, 4.0)$, and $(3.0, 5.0)$.
First, the collisional components of turbulent energy transfer $Q^{\rm turb}_{aC}(a=e, i)$ are found to be negligible under the conditions.
Parallel heating ($Q^{\rm turb}_{a \parallel} > 0$ for $a = e, i$) and perpendicular cooling ($Q^{\rm turb}_{a B} < 0$ for $a = e, i$) are observed across the wavenumber space, except for the zonal mode.
Comparison between Figs.~\ref{fig:partQspectrum_mixedITGTEM} (a, b) and (e, f) shows that the zonal contribution to $Q^{\rm turb}_{a} (a=e, i)$ in ITG-dominated turbulence is greater than that in TEM-dominated turbulence.
Correspondingly, the nonlinear mode-coupling terms $Q^{\rm turb}_{a\psi\bm{k}}(a=e,i)$ in Eq.~(\ref{eq:psi_heating}) show greater contributions in the unstable wavenumber region ($0.1\le k_y\rho_{ti}\le0.3$) for ITG turbulence than for TEM turbulence.
Although these nonlinear terms do not directly affect the net energy exchange, they influence the redistribution of energy from the peak region to both the zonal modes and higher wavenumber modes.
The effect of $Q^{\rm turb}_{a\psi\bm{k}}(a=e, i)$ appears significantly in the case of $(R_0/L_{Te}, R_0/L_{Ti})=(4.0, 4.0)$ as shown in Figs.~\ref{fig:partQspectrum_mixedITGTEM} (c) and (d).
Interestingly, although the total turbulent energy exchange for ions is negative, $Q^{\rm turb}_{i}<0$, as in ITG turbulence, the sign of $Q^{\rm turb}_{i\bm{k}}$ in the peak wavenumber mode ($k_y\rho_{ti}=0.20$) is positive, resembling that of TEM turbulence.
This phenomenon is attributed to the inequality $|Q^{\rm turb}_{a\parallel\bm{k}}+Q^{\rm turb}_{aB\bm{k}}|<|Q^{\rm turb}_{a\psi\bm{k}}|(a=e, i)$.
In other words, in turbulence where ITG and TEM are of comparable strength, the contributions from the zonal mode and higher wavenumber regions can exceed those from the unstable wavenumber region.
Figure~\ref{fig:partQ_LTi_mixedITGTEM} shows the parallel heating and perpendicular cooling for electrons and ions as functions of $(R_0/L_{Te}, R_0/L_{Ti})$.
The sum of the parallel and perpendicular components, as well as the total turbulent energy transfer, are also shown in Fig.~\ref{fig:partQ_LTi_mixedITGTEM}.
It is confirmed that turbulent energy transfer $Q^{\rm turb}_{a} (a=e, i)$ is almost equal to the sum of the parallel heating and perpendicular cooling components $Q^{\rm turb}_{a\parallel}+Q^{\rm turb}_{aB}(a=e, i)$.
When the TEM mode is dominant, the absolute value of parallel heating and perpendicular cooling terms for electrons $|Q^{\rm turb}_{e\parallel}|, |Q^{\rm turb}_{eB}|$, as well as parallel heating for ions $|Q^{\rm turb}_{i\parallel}|$ are large and the perpendicular cooling for ions $|Q^{\rm turb}_{iB}|$ is small.
In contrast, when the ITG mode is dominant, the parallel heating and perpendicular cooling of ions and parallel heating of electrons, $|Q^{\rm turb}_{i\parallel}|, |Q^{\rm turb}_{iB}|, |Q^{\rm turb}_{e\parallel}|$ increase, and the perpendicular cooling for electrons $|Q^{\rm turb}_{eB}|$ is small. 
%

%
Having verified the correlation between energy flux and energy exchange defined in Eq.~(\ref{eq:Qturb_EFlux}), we now examine the validity of the hypothesis presented in Eqs.~(\ref{eq:perp_heating_ITG}) and (\ref{eq:Qpara_Eflux}).
Figure~\ref{fig:partQ_EF_mixedITGTEM} shows the parallel heating and perpendicular cooling components of energy exchange for electrons and ions as functions of $(\mathcal{E}^{\mathrm{turb}}_e+\mathcal{E}^{\mathrm{turb}}_i)/2R_0$, and $\mathcal{E}^{\mathrm{turb}}_a/R_0$, respectively.
It is found that the perpendicular coolings for electrons and ions are proportional to each energy flux.
In particular, they appear to be on the same straight line passing through the origin, and the slope is approximately -1.
Then, the validity of Eq.~(\ref{eq:perp_heating_ITG}) is demonstrated.
Although the parallel heating exhibits greater scatter than the perpendicular heating, it maintains a proportional relationship, with a slope of unity, to the average energy flux of each particle species divided by the major radius.
From these results, it is clarified that the perpendicular heating terms are primarily determined by the energy flux of their respective particle species.

\begin{figure*}[tbp]
    \includegraphics[keepaspectratio, scale=0.47]{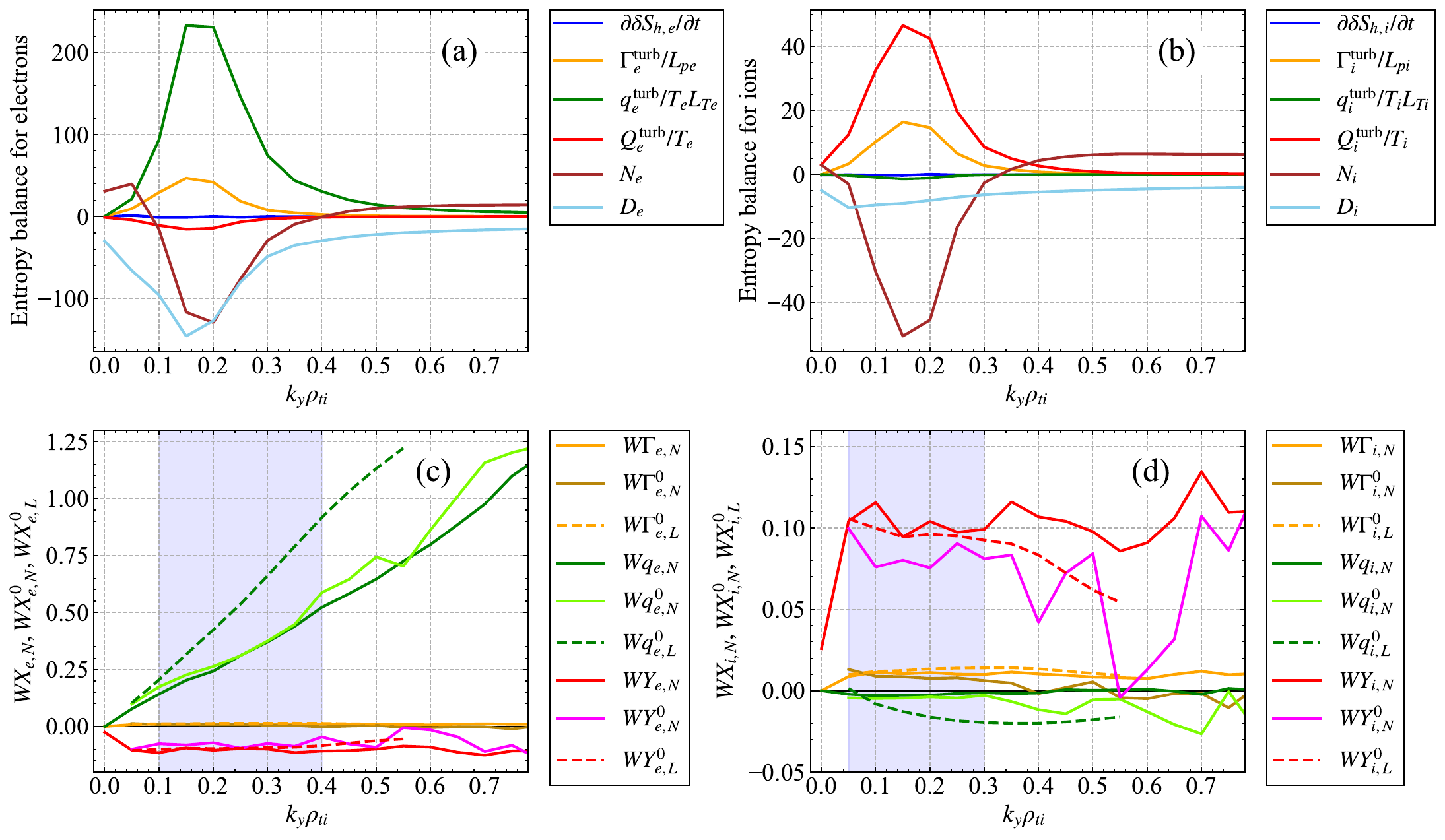}
    \caption{The wavenumber spectra of terms in the entropy balance equation in Eq.~(\ref{eq:EBequation_wavenumber}) for electrons (a) and ions (b) in the case of $(R_0/L_{Te}, R_0/L_{Ti}) =(7.0, 1.0)$. 
    The spectra are given as functions of $k_y \rho_{ti}$ obtained by summing over $k_x$. 
    They are evaluated in the steady state of turbulence obtained by nonlinear simulation. 
    Nonlinear and quasilinear weights are shown for electrons (c) and ions (d).
    The nonlinear entropy transfer terms $N_{ek_y}$ and $N_{ik_y}$ are negative in the wavenumber regions colored in blue. 
    Dashed lines represent quasilinear weights obtained by linearly unstable modes with $k_x  = 0$.
}
    \label{fig:QuasilinearWeights_pureTEM}
\end{figure*}

\subsection{Quasilinear model for ITG-TEM turbulence}\label{subsec:3D}

\begin{figure}[tbp]
    \includegraphics[keepaspectratio, scale=0.33]{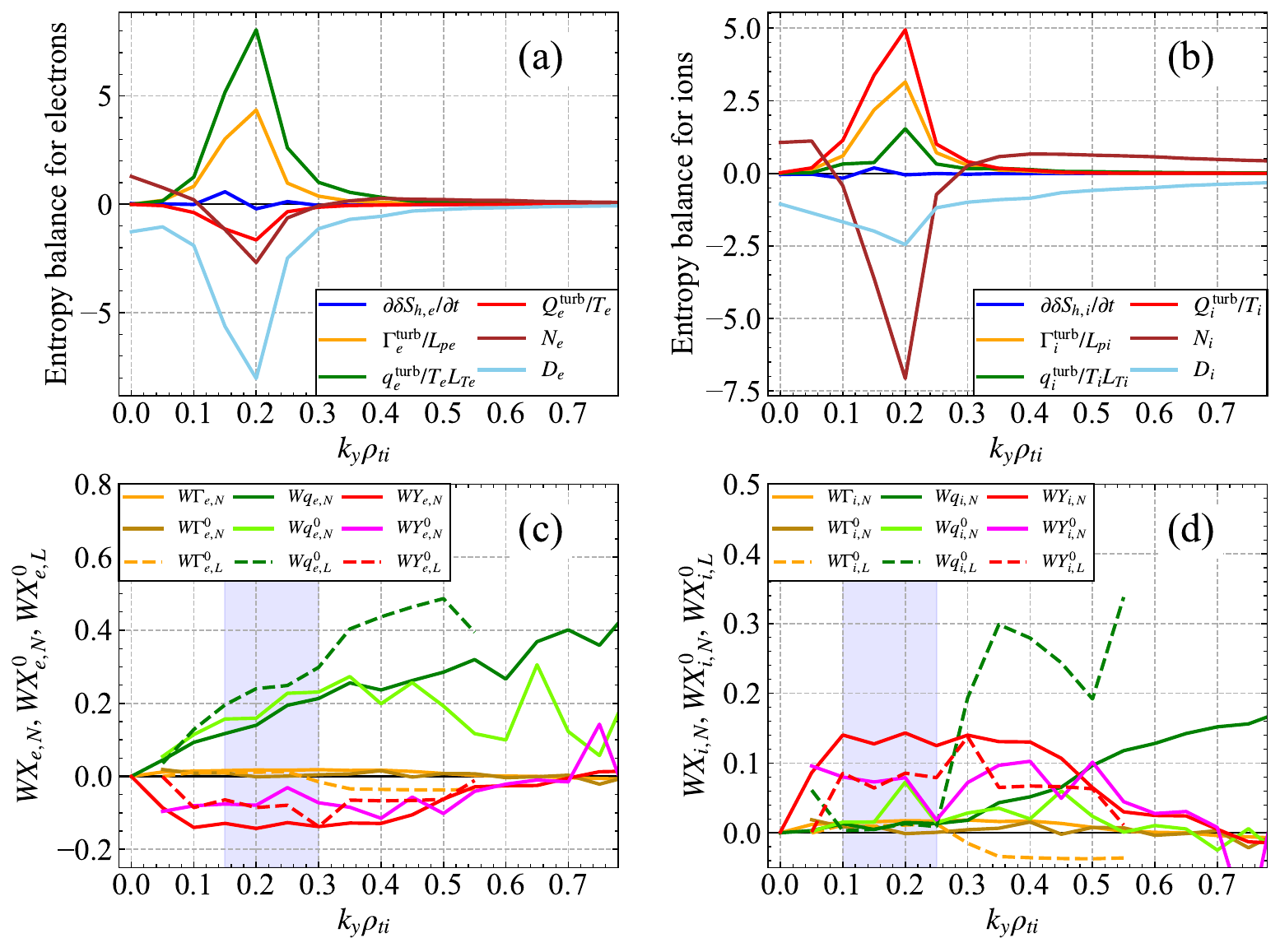}
    \caption{
    The wavenumber spectra of terms in the entropy balance equation in Eq.~(\ref{eq:EBequation_wavenumber}) for electrons (a) and ions (b) in the case of $(R_0/L_{Te}, R_0/L_{Ti}) =(5.0, 3.0)$. 
    Nonlinear and quasilinear weights are shown for electrons (c) and ions (d).
    The nonlinear entropy transfer terms $N_{ek_y}$ and $N_{ik_y}$ are negative in the wavenumber regions colored in blue. 
}\label{fig:QuasilinearWeights_5.3}
\end{figure}

\begin{figure}[tbp]
    \includegraphics[keepaspectratio, scale=0.33]{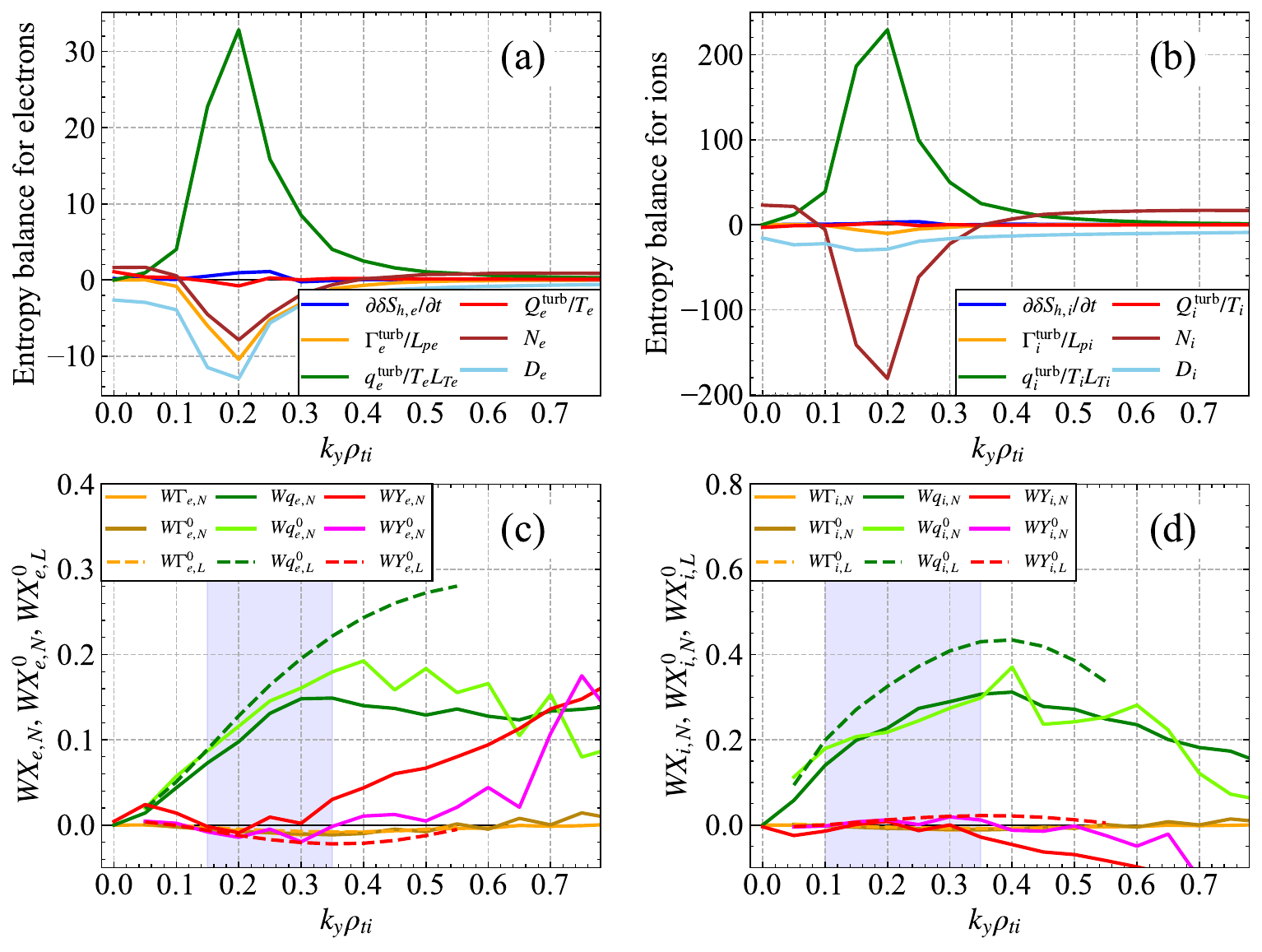}
    \caption{
    The wavenumber spectra of terms in the entropy balance equation in Eq.~(\ref{eq:EBequation_wavenumber}) for electrons (a) and ions (b) in the case of $(R_0/L_{Te}, R_0/L_{Ti}) =(4.0, 4.0)$. 
    Nonlinear and quasilinear weights are shown for electrons (c) and ions (d).
    The nonlinear entropy transfer terms $N_{ek_y}$ and $N_{ik_y}$ are negative in the wavenumber regions colored in blue. 
}\label{fig:QuasilinearWeights_4.4}
\end{figure}

\begin{figure}[tbp]
    \includegraphics[keepaspectratio, scale=0.33]{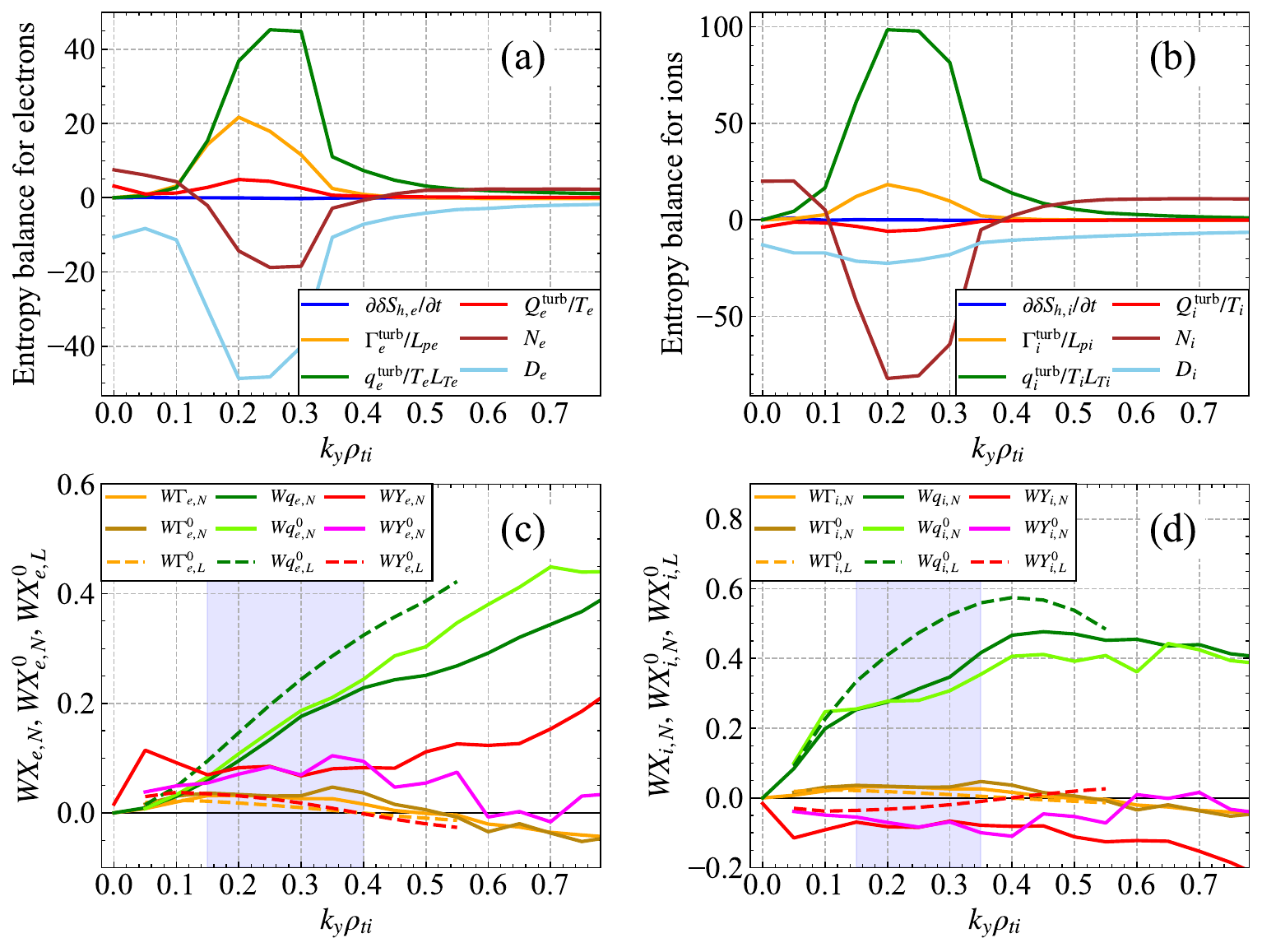}
    \caption{
    The wavenumber spectra of terms in the entropy balance equation in Eq.~(\ref{eq:EBequation_wavenumber}) for electrons (a) and ions (b) in the case of DIII-D shot $128913$ $(T_e/T_i, R_0/L_{na}, R_0/L_{Te}, R_0/L_{Ti}) =(1.2, 3.0, 6.5, 5.0)$. 
    Nonlinear and quasilinear weights are shown for electrons (c) and ions (d).
    The nonlinear entropy transfer terms $N_{ek_y}$ and $N_{ik_y}$ are negative in the wavenumber regions colored in blue. 
}\label{fig:QuasilinearWeights_Candy}
\end{figure}

Finally, we investigate the applicability of quasilinear model for turbulent energy exchange in ITG-TEM turbulence.
Figures~\ref{fig:QuasilinearWeights_pureTEM} (a) and (b) show the wavenumber spectra of terms in the entropy balance equation in Eq.~(\ref{eq:EBequation_wavenumber}) in pure TEM turbulence.
Entropy produced by particle and heat fluxes for electrons is partially transferred to ions.
Additionally, entropy is redistributed among wavenumber modes via nonlinear interactions $N_{e\bm{k}}$ and $N_{i\bm{k}}$, which transfer entropy from unstable modes to zonal and high-wavenumber modes.
Although the entropy transfer from unstable to stable modes within the same $k_y$ region may also occur under this condition \cite{Terry}, such an analysis has not been conducted in this study.
Meanwhile, entropy is dissipated through collisions in a wide wavenumber range, which maintains the overall entropy balance.
This mechanism is similar to the situation in ITG turbulence, except that the roles of electrons and ions are reversed.
Figures~\ref{fig:QuasilinearWeights_pureTEM} (c) and (d) show the wavenumber spectra of nonlinear and quasilinear weights in pure TEM turbulence.
The nonlinear weights calculated at $k_x\rho_{ti}=0$, $WX^0_{a,N} (X_a=\Gamma_a, q_a, Y_a$ for $a=e, i)$ are found to agree well with those obtained by summing over $k_x$ modes, $WX_{a,N}$, respectively, which indicates that the result from linear simulations for $k_x=0$ mode is appropriate for evaluating quasilinear weights.
It is found that quasilinear weights for energy exchange sufficiently agree with their nonlinear weights 
 within an error margin of $15 \%$ or less in the colored wavenumber region where $N_{e\bm{k}}<0$ and $N_{i\bm{k}}<0$ , while those for particle and heat fluxes are within a factor of two except for ion heat flux.
Furthermore, the discrepancy between quasilinear and nonlinear weights becomes more significant in higher wavenumber modes, beyond the colored wavenumber region.
Since the colored wavenumber region accounts for $91\%$ of the total energy exchange, the validity of the quasilinear model for energy exchange in pure TEM turbulence is confirmed.
Next, the feasibility of the quasilinear model for energy exchange in ITG-TEM turbulence is examined.
Figures~\ref{fig:QuasilinearWeights_5.3} and \ref{fig:QuasilinearWeights_4.4} show the wavenumber spectra of terms in the entropy balance equation along with nonlinear and quasilinear weights for $(R_0/L_{Te}, R_0/L_{Ti})=(5.0, 3.0)$ and $(4.0, 4.0)$, respectively.
These two cases are regarded as the results of turbulence where both ITG and TEM coexist.
In the case of $(R_0/L_{Te}, R_0/L_{Ti})=(5.0, 3.0)$, entropy is largely produced by the electron particle and heat fluxes, and entropy transfer from electrons to ions is the main source of entropy production for ions.
This situation is similar to pure TEM turbulence.
For both electrons and ions $(a=e, i)$, the values of nonlinear weights for energy transfer calculated at $k_x\rho_{ti}=0$, $WY^0_{a,N}$, are approximately half the values of $WY_{a,N}$ in the wavenumber regions where $N_{a\bm{k}}<0$.
The quasilinear weights for energy exchange $WY^0_{a,L}$ agree with $WY^0_{a,N}$ in the wavenumber regions.
Although the discrepancy between the nonlinear and quasilinear weights for energy transfer is larger than that observed in pure ITG or TEM turbulence, the level of agreement remains comparable to that of particle and heat fluxes within a factor of two.
A similar trend is observed for all parameter sets except for $(R_0/L_{Te}, R_0/L_{Ti}) = (4.0, 4.0)$.
In this case, the heat fluxes of both electrons and ions serve as the primary sources of entropy production associated with the perturbed distribution function, while the particle fluxes act to reduce the perturbed entropy.
Since the contribution from energy exchange to the entropy balance is small, each particle species maintains its own entropy balance through its respective fluxes and collisional dissipation.
The quasilinear weights for particle and heat fluxes agree with the nonlinear results, whereas the weight for energy exchange shows discrepancies in both magnitude and sign in certain regions.
This is because, as shown in Fig.~\ref{fig:partQspectrum_mixedITGTEM}, the sign of the energy exchange $Q^{\rm turb}_{a\bm{k}}$ changes for each wavenumber mode due to the nonlinear interaction term, $Q^{\rm turb}_{a\psi\bm{k}}$. 
As a result, the linear simulation for $(R_0/L_{Te}, R_0/L_{Ti}) = (4.0, 4.0)$ fails to provide the quasilinear weight that reasonably reproduces the nonlinear weight for energy transfer.
Here, using the quasilinear weights, quasilinear fluxes and energy exchange are defined as,
\begin{equation}
X_{a,QL}=\sum_{0<k_y\rho_{ti}\le0.55}\left\{WX^0_{a, L}(k_y)\left(\sum_{k_x}\langle \langle |\phi_{\bm{k}, N}|^2 \rangle\rangle\right)(k_y)\right\}. \label{eq:QuasilinearFluxes}
\end{equation}
The quasilinear results $X_{a,QL}(X_a=\Gamma_a, q_a, Y_a$ for $a=e, i)$ here represent fluxes and energy exchange predicted using quasilinear weights obtained from linear simulations, together with the $k_y$ spectrum of the perturbed electrostatic potential obtained from nonlinear simulations.
The quasilinear fluxes and energy exchange are shown in Fig.~\ref{fig:PF_HF_EE_LTi_ITGTEM}.
It can be seen that the quasilinear model tends to overestimate the fluxes while underestimating the energy exchange.
Additionally, it is confirmed that the energy transfer in ITG-TEM turbulence can be predicted by the quasilinear model as well as particle and heat fluxes, except for $(R_0/L_{Te}, R_0/L_{Ti}) = (4.0, 4.0)$.
For the case of $(R_0/L_{Te}, R_0/L_{Ti}) = (4.0, 4.0)$ [see also Fig.~\ref{fig:QuasilinearWeights_4.4}], the quasilinear model fails to predict even the correct sign of the energy transfer calculated by the nonlinear simulation.
In this case, both ITG and TEM modes coexist and contribute in opposite directions to the energy exchange, largely canceling each other out and resulting in a small net turbulent energy transfer.
Nevertheless, the quasilinear weight is calculated using only the most unstable mode.
Instead of directly applying the quasilinear model to energy transfer, we now attempt to predict it using a combination of the quasilinear model and Eq.~(\ref{eq:Qturb_EFlux}), expressed as
\begin{equation}
    \Delta EF_{QL}=\frac{\mathcal{E}_{e, QL}-\mathcal{E}_{i, QL}}{2R_0},
    \label{eq:DeltaEF}
\end{equation}
where $\mathcal{E}_{a, QL}=q_{a, QL}+5T_a\Gamma_{a, QL}/2$.
The results calculated by Eq.(\ref{eq:DeltaEF}), along with $\Delta EF$ computed from the nonlinear simulation, are shown in Fig.\ref{fig:PF_HF_EE_LTi_ITGTEM}.
The energy flux difference between electrons and ions from the nonlinear simulation $\Delta EF$ shows good agreement with the turbulent energy exchange, including the sign.
These results suggest that the model based on Eq.~(\ref{eq:Qturb_EFlux}) or (\ref{eq: delta EF}) has the potential to predict turbulent energy exchange effectively.
It is found that the signs of energy exchange obtained by Eq.~(\ref{eq:DeltaEF}) match those from nonlinear simulations for all parameter sets, including the critical case of $(R_0/L_{Te}, R_0/L_{Ti}) = (4.0, 4.0)$.
On the other hand, because the quasilinear weights for particle and heat fluxes tend to be overestimated, the magnitudes of energy exchange obtained from Eq.~(\ref{eq:DeltaEF}) are roughly twice as large as the nonlinear simulation results.
This discrepancy can be mitigated by employing a model that more accurately predicts energy fluxes or by applying an appropriate correction factor.
However, the signs do not necessarily match perfectly even with Eq.~(\ref{eq:DeltaEF}).
In such cases, the magnitude of the turbulent energy exchange is very small and can therefore be neglected.
Hence, this sign inconsistency is acceptable.
In addition, the results of linear and nonlinear simulations based on the experimental parameters of DIII-D shot~128913\cite{White} ($T_e/T_i=1.2, R_0/L_{Te}=6.5, R_0/L_{Ti}=5.0, R_0/L_{na}=3.0$, other parameters are same as in Tab.\ref{tab:Plasma and field parameters}) are presented in Fig.~\ref{fig:QuasilinearWeights_Candy}.
As shown in Fig.~\ref{fig:QuasilinearWeights_Candy}, transport fluxes for both electrons and ions contribute to the generation of perturbed entropy.
The direction of entropy and energy transfer is from ions to electrons, which is consistent with the characteristic of ITG-dominated turbulence.
The quasilinear weights for the particle and heat fluxes show reasonable agreement with the corresponding nonlinear weights.
Although the quasilinear weights for energy exchange are less than half of the nonlinear weights, their signs are consistent with the nonlinear results.
The ratios of the quasilinear fluxes and energy exchange calculated by Eq.~(\ref{eq:QuasilinearFluxes}) to the nonlinear simulation results are 57\%, 133\%, 136\%, and 27\% for the particle fluxes, electron and heat fluxes, and energy transfer, respectively.
The energy transfer calculated from Eq.~(\ref{eq:DeltaEF}) is 87\% of the nonlinear result, which is better than the result predicted by Eq.~(\ref{eq:QuasilinearFluxes}).
Consequently, although the accuracy of the quasilinear model for energy exchange is lower than that for particle and heat fluxes, these results demonstrate that it can still provide a reasonable estimate of energy transfer in parameter regimes relevant to experimental conditions.

\section{Conclusions and Discussion}\label{sec:4}
In this study, the turbulent energy transfer in pure TEM and mixed ITG-TEM turbulence in tokamak plasmas is investigated.
It is confirmed that pure TEM turbulence transfers energy from electrons to ions, while pure ITG turbulence transfers energy from ions to electrons. 
The wavenumber spectral analysis clarifies that energy is transferred from electrons to the perturbed electrostatic potential via the cooling of electrons in the $\nabla B$-curvature drift motion and from the potential to ions via heating of ions streaming along the field line, which are the main mechanisms of turbulent energy exchange in pure TEM turbulence.
From the viewpoint of entropy balance, the particle and heat fluxes of electrons generate large entropy associated with the perturbed distribution function, and energy exchange plays a role in transferring part of the entropy to ions.
Thus, TEM turbulence is expected to facilitate the transfer of energy from alpha-heated electrons to ions, contributing to more efficient ion heating in fusion plasmas.
The investigation of mixed ITG-TEM turbulence reveals that the direction of energy transfer reverses as the most unstable mode transitions between ITG and TEM.
It is also confirmed that the sign of energy exchange aligns with that of the difference in entropy production.
Therefore, this result supports the conjecture that energy is transferred by turbulence from particle species with large entropy production due to turbulent transport to the other species.
The signs of parallel heating and perpendicular cooling remain unchanged whether ITG or TEM is dominant, and the relative magnitude of these two effects determines the magnitude and direction of total turbulent energy transfer.
In turbulence where ITG and TEM instabilities are of comparable strength, the contribution of the sum of parallel heating and perpendicular cooling in the unstable wavenumber region can become smaller than that of the nonlinear interaction term.
Then, the sign of energy exchange for unstable wavenumber modes may differ from that of total energy transfer, and the contribution of the zonal mode and high wavenumber modes should not be negligible.
It is also found that the perpendicular cooling of each particle species is strongly correlated with its own energy flux.
In order to examine the validity of the quasilinear model in ITG-TEM turbulence, quasilinear weights for turbulent fluxes and energy exchange are also investigated.
In the case of turbulence where the unstable mode is significantly biased toward ITG or TEM, the quasilinear weight for energy exchange agrees with the nonlinear weight within an error margin of 20\%, which is better than that for particle and heat fluxes.
On the other hand, when both ITG and TEM instabilities contribute comparably to the turbulence, the accuracy of quasilinear predictions for turbulent fluxes remains almost unchanged, whereas that for energy exchange deteriorates.
In addition, it is revealed that when the wavenumber spectrum of energy transfer contains both positive and negative contributions, the quasilinear model may predict the opposite sign of energy exchange compared to the result obtained from nonlinear simulations.
In such cases, it is worthwhile to examine quasilinear weights by taking into account not only the most unstable mode but also the subdominant unstable modes\cite{Staebler2021, Staebler2024, Merz}.
Furthermore, we propose a method to predict turbulent energy transfer based on the difference in energy flux between electrons and ions.
This approach successfully reproduces the direction of energy transfer, even in a case where the quasilinear model fails.
This model is constructed from the correlation between the energy flux and the perpendicular cooling, which proves useful for estimating nonlinear simulation results of energy transfer.
By employing models that can predict energy fluxes in greater detail, or by introducing appropriate correction coefficients, it is expected that turbulent energy transfer in ITG–TEM turbulence can be predicted more accurately, thereby contributing to more reliable profile predictions in operations for future fusion reactors.

\section*{acknowledgements}
The present study is supported in part by the JSPS Grants-in-Aid for Scientific Research Grant No.~24K07000 and in part by the NIFS Collaborative Research Program NIFS23KIPT009. 
This work is also supported by JST SPRING, Grant Number JPMJSP2108.
Simulations in this work were performed on “Plasma Simulator” (NEC SX-Aurora TSUBASA) of NIFS with the support and under the auspices of the NIFS Collaboration Research program (NIFS24KISM007). 

\section*{author declarations}
\subsection*{Conflict of Interest}
The authors have no conflicts to disclose.

\subsection*{Author Contributions}
\noindent
{\bf Tetsuji Kato}: 
Conceptualization (equal);
Data curation (lead);
Formal analysis (equal);
Investigation (lead);
Methodology (equal);
Writing – original draft (lead);
Writing – review \& editing (lead).
{\bf Hideo Sugama}: 
Conceptualization (equal);
Formal analysis (equal);
Investigation (supporting);
Methodology (equal);
Supervision (lead);
Writing – original draft (supporting);
Writing – review \& editing (supporting).
{\bf Tomohiko Watanabe}: 
Methodology (equal);
Writing – review \& editing (supporting). 

\section*{Data availability}
The data that support the findings of this study are available from the corresponding authors upon reasonable request.

\section*{references}


\begin{references}


\bibitem{Kato2024}
T. Kato, H. Sugama, T.-H. Watanabe, and M. Nunami,   
Phys.\ Plasmas {\bf 31}, 062510 (2024).

\bibitem{Sugama1996}
H. Sugama, M. Okamoto, W. Horton, and M. Wakatani,   
Phys.\ Plasmas {\bf 3}, 2379 (1996).

\bibitem{Sugama2009}
H. Sugama, T.-H. Watanabe, and M. Nunami,
Phys.\ Plasmas {\bf 16}, 112503 (2009). 

\bibitem{Candy}
J. Candy, 
Phys.\ Plasmas {\bf 20}, 082503 (2013). 

\bibitem{Manheimer}
W. M. Manheimer, E. Ott, and W. M. Tang,
Phys.\ Fluids {\bf 20}, 808 (1977).

\bibitem{Waltz1997}
R. E. Waltz, G. W. Staebler, W. Dorland, G. W. Hammett, M. Kotschenreuther, and J. A. Konings, 
Phys.\ Physics {\bf 1}, 2482 (1997).

\bibitem{Waltz}
R. E. Waltz, and G. M. Staebler,
Phys.\ Plasmas {\bf 15}, 014505 (2008). 




\bibitem{Antonsen}
T. M. Antonsen, Jr.\ and B. Lane, 
Phys.\ Fluids {\bf 23}, 1205 (1980).

\bibitem{CTB}
P.J. Catto, W.M. Tang and D.E. Baldwin, Plasma Phys.
{\bf 23}, 639 (1981).

\bibitem{F-C}
E. A. Frieman and L. Chen, 
Phys.\ Fluids {\bf 25}, 502 (1982).

\bibitem{Pueschel}
M. J. Pueschel and F. Jenko,
Phys.\ Plasmas {\bf 17}, 062307 (2010).


\bibitem{Dimits}
A. M. Dimits, G. Bateman, M. A. Beer, B. I. Cohen, W. Dorland, G. W. Hammett, C. Kim, J. E. Kinsey, M. Kotschenreuther, A. H. Kritz, L. L. Lao, J. Mandrekas, W. M. Nevins, S. E. Parker, A. J. Redd, D. E. Shumaker, R. Sydora, and J. Weiland, 
Phys.\ Plasmas {\bf 7}, 969 (2000).


\bibitem{Krommes2012}
J.A. Krommes, 
Ann.\ Rev.\ Fluid Mech.\ {\bf 44} 175 (2012). 

\bibitem{Garbet2010}
X. Garbet, Y. Idomura, L. Villard, and T.-H. Watanabe, 
Nucl.\ Fusion {\bf 50} 043002 (2010). 

\bibitem{Idomura2006}
Y. Idomura, T.-H. Watanabe, and H. Sugama, 
Comptes Rendus Physique {\bf 7}, 650 (2006). 


\bibitem{Horton}
W. Horton, 
{\it Turbulent Transport in Magnetized Plasmas}, 
2nd edition (World Scientific, Singapore, 2018), Chap.12. 



\bibitem{Casati}
A. Casati, C. Bourdelle, X. Garbet, F. Imbeaux, J. Candy, F. Clairet, G. Dif-Pradalier, G. Falchetto, T. Gerbaud, V. Grandgirard, Ö.D. Gürcan, P. Hennequin, J. Kinsey, M. Ottaviani, R. Sabot, Y. Sarazin, L. Vermare, and R.E. Waltz,
Nucl.\ Fusion {\bf 49}, 085012 (2009). 

\bibitem{Citrin}
J. Citrin, C. Bourdelle, P. Cottier, D. F. Escande, Ö. D. Gürcan, D. R. Hatch, G. M. D. Hogeweij, F. Jenko, and M. J. Pueschel, 
Phys.\ Plasmas {\bf 19} 062305 (2012).

\bibitem{Citrin2017}
J. Citrin, H Arnichand, J Bernardo, C Bourdelle, X Garbet, F Jenko, S Hacquin, M J Pueschel, and R Sabot, 
Plasma Phys.\ Control.\ Fusion {\bf 59} 064010(2017).

\bibitem{Bourdelle}
C. Bourdelle, J. Citrin, B. Baiocchi, A. Casati, P. Cottier, X. Garbet, F. Imbeaux, and JET Contributors, 
Plasma Phys.\ Control.\ Fusion {\bf 58} 014036(2016). 

\bibitem{Staebler2021}
G. M. Staebler, E. A. Belli, J. Candy, J. E. Kinsey, H. Dudding, and B. Patel, 
Nucl.\ Fusion {\bf 61}, 116007 (2021). 

\bibitem{Staebler2024}
G. Staebler, C. Bourdelle, J. Citrin, and R. Waltz
Nucl.\ Fusion {\bf 64}, 103001 (2024). 

\bibitem{Merz}
F. Merz and F. Jenko
Nucl.\ Fusion {\bf 50} 054005 (2010).

\bibitem{Narita}
E. Narita, M. Honda, M. Nakata, M. Yoshida, and N. Hayashi,
Nucl.\ Fusion {\bf 61}, 116041 (2021). 

\bibitem{Toda}
S. Toda, M. Nunami, and H. Sugama, 
Plasma Phys.\ Control.\ Fusion {\bf 64}, 085001 (2022). 

\bibitem{Parker}
S. E. Parker, C. S. Haubrich, S. Tirkas, Q. Cai, and Y. Chen,
Plasma {\bf 6},  611(2023). 


\bibitem{Barnes}
M. Barnes, P. Abiuso, and W. Dorland, J. Plasma Phys.\ {\bf 84}, 905840306 (2018).

\bibitem{GKV}
T.-H. Watanabe and H. Sugama, Nucl.\ Fusion {\bf 46}, 24 (2006).

\bibitem{Beer}
M. A. Beer, S. C. Cowley, and G. W. Hammett,
Phys.\ Plasmas {\bf 2}, 2687 (1995). 

\bibitem{Hinton1976}
F. L. Hinton and R. D. Hazeltine,
Rev.\ Mod.\ Phys. {\bf 48}, 239 (1976).

\bibitem{Nakata}
M. Nakata, M. Nunami, H. Sugama, and T.-H. Watanabe,
Phys.\ Rev.\ Lett.\ {\bf 118}, 165002 (2017).

\bibitem{Lee}
W.W Lee,
J. Comp.\ Phys.\ {\bf 72}, 243 (1987).

\bibitem{Lenard}
A. Lenard and Ira B. Bernstein,
Phys.\ Rev.\ {\bf 112}, 1456 (1958).

\bibitem{Ball}
J. Ball, S. Brunner, and C.J. Ajay,
J.\ Plasma\ Phys.\ {\bf 86}, 905860207 (2020).

\bibitem{Terry}
P. W. Terry, K. D. Makwana, M. J. Pueschel, D. R. Hatch, F. Jenko, and F. Merz,
Phys.\ Plasmas {\bf 21}, 122303 (2014).

\bibitem{White}
A. E. White, L. Schmitz, G. R. McKee, C. Holland, W. A. Peebles, T. A. Carter, M. W. Shafer, M. E. Austin, K. H. Burrell, J. Candy, J. C. DeBoo, E. J. Doyle, M. A. Makowski, R. Prater, T. L. Rhodes, G. M. Staebler, G. R. Tynan, R. E. Waltz, and G. Wang,
Phys.\ Plasmas {\bf 15}, 056116 (2008).

\end{references}
\end{document}